\documentclass[aps,twocolumn,groupedaddress]{revtex4}
\usepackage{epsfig}
\usepackage{graphicx}
\usepackage[T1]{fontenc}
\usepackage{ae}
\usepackage{color}
\usepackage[latin1]{inputenc}
\usepackage{amssymb,amsbsy,amsmath}
\usepackage{bbm}

\newtheorem{ass}{Assertion}

\begin{document}

%\begin{center}\today\end{center}

%%%%%%%%%%%%%%%%%%%%%%%%%%%%%%%%%%%%%%%%%%%%%%%%%%%%%%%%%%%%%%%%%%%%%%%%%%%%%

\title[RPL]
{The Floquet theory of the two level system revisited}
\author{Heinz-J\"urgen Schmidt$^1$
\footnote[1]{Correspondence should be addressed to hschmidt@uos.de}
}
\address{$^1$Universit\"at Osnabr\"uck, Fachbereich Physik,
Barbarastr. 7, D - 49069 Osnabr\"uck, Germany}

%\tableofcontents

\begin{abstract}
We reconsider the periodically driven two level system and especially the Rabi problem with linear polarization.
The Floquet theory of this problem can be reduced to its classical limit, i.~e.~, to the investigation of periodic
solutions of the classical Hamiltonian equations of motion in the Bloch sphere. The quasienergy is essentially the action
integral over one period and the resonance condition due to J.~H.~Shirley is shown to be equivalent to the vanishing
of the time average of a certain component of the classical solution. This geometrical approach is applied to
obtain analytical approximations to physical quantities of the Rabi problem with linear polarization
as well as asymptotic formulas for various limit cases.
\end{abstract}

\maketitle

%%%%%%%%%%%%%%%%%%%%%%%%%%%%%%%%%%%%%%%%%%%%%%%%%%%%%%%%%%%%%%%%%%%%%%%%%%%%%%%%%%%%%%%%%%%%%%%%%%%%%%%%%%%%%%%%%%%%%%%%%%%%%%%
\section{Introduction}\label{sec:Intro}
%%%%%%%%%%%%%%%%%%%%%%%%%%%%%%%%%%%%%%%%%%%%%%%%%%%%%%%%%%%%%%%%%%%%%%%%%%%%%%%%%%%%%%%%%%%%%%%%%%%%%%%%%%%%%%%%%%%%%%%%%%%%%%%

Many physical experiments can be described as the interaction of a small quantum system with electromagnetic radiation.
If one tries to theoretically simplify this situation as far as possible one arrives at a two level system (TLS) interacting
with a classical periodic radiation field. The special case of a constant magnetic field into, say, $z$-direction plus
a circularly polarized field in the $x-y$-plane has been solved eight decades ago by I.~I.~Rabi \cite{R37} and found its way
into many textbooks. This case will be referred to as the RPC.
Shortly after, F.~Bloch and A.~Siegert \cite{BS40} considered the analogous problem of a linearly polarized magnetic field
orthogonal to the direction of the constant field (henceforward called RPL) and suggested the so-called rotating wave approximation.
Moreover, they investigated the shift of the resonance frequencies due to the approximation error of the rotating wave approximation,
since then called ''Bloch-Siegert shift".

In the following decades it has been realized \cite{AT55} \cite{S65} that the underlying mathematical problem is an instance of the Floquet theory
that deals with linear differential matrix equations with periodic coefficients. Accordingly analytical approximations for
its solutions have been devised that have been the basis for subsequent research. Especially the seminal paper of J.~H.~Shirley \cite{S65} has
until now been cited more than $11\,000$ times. Among the numerous applications of the theory of periodically driven two level systems
are nuclear magnetic resonance \cite{R38},  ac-driven quantum dots \cite{WT10},  Josephson qubit circuits \cite{YN11}, and
coherent destruction of tunneling \cite{MZ16}.
On the theoretical level the methods of solving the RPL and related problems have been gradually refined
and include power series approximations for Bloch-Siegert shifts \cite{HPS73} \cite{AB74},
perturbation theory and/or various limit cases  \cite{LP92} -- \cite{WY07} and the hybridized rotating wave approximation \cite{YLZ15}.
Special analytical solutions of the general Floquet problem of TLS can be generated by an inverse method \cite{GDG10} -- \cite{MN14}.

There exists even an analytical solution \cite{XH10}
of the RPL and its generalization to the Rabi problem where the angle between the constant field and the periodic field is arbitrary.
This solution bears on a transformation of the Schr\"odinger equation into a confluent Heun differential equation.
A similar approach has previously been applied to the TLS subject to
a magnetic pulse \cite{JR10}. However, the analytic solution is achieved by gluing together three different solutions and does not
yield explicit solutions for the quasienergy or for resonance curves. The ongoing research on Heun functions, see, e.~g.~, \cite{C04} \cite{EF13},
might facilitate the physical interpretation of these analytical solutions in the future.
Summarizing, the problem is far from being completely solved
and it appears still worthwhile to further investigate the general
Floquet problem of the TLS and to look for more analytical approximations of the RPL.

In this paper we will suggest an approach to the Floquet problem of the TLS via its well-known classical limit, see, e.~g.~, \cite{AW05}.
It turns out that, surely not in general, but for this particular problem, the classical limit is already equivalent to the quantum problem.
More precisely, we will show that to each periodic solution of the classical equation of motion there exists a Floquet solution
of the original Schr\"odinger equation that can be explicitly calculated via integrations. Especially, the quasienergy is essentially given by the action
integral over one period of the classical solution. This is reminiscent of the semiclassical Floquet theory developed in \cite{BH91}.
In the special case of the Rabi problem with elliptical polarization our approach yields the result that Shirley's resonance condition
is equivalent to the vanishing of the time average of the component of the classical periodic solution into the direction of the constant
magnetic field. When applied to the RPL our approach suggests to calculate truncated Fourier series for the classical solution and to obtain
from this the quasienergy by the recipe sketched above. For the various limit cases of the RPL there exist also classical versions that
will be analyzed and evaluated in order to obtain asymptotic formulas for the quasienergy.

The structure of the paper is the following. We have three main parts, Generalities, Resonances and Analytical Approximations,
that refer to problems of decreasing generality: The general two level system (TLS) with a periodic Hamiltonian,  the Rabi problem
with elliptic polarization (RPE) and the Rabi problem with linear polarization (RPL). Moreover, we have four subsections
\ref{sec:RPCI},\, \ref{sec:RPCII},\,\ref{sec:RPCIII}, and \ref{sec:EX},
where the explicitly solvable case of the Rabi problem with circular polarization (RPC) and another solvable toy example
is used to illustrate certain results of the first two main parts.

In subsection \ref{sec:GenSU2} we start with a short account of the well-known Floquet theory of TLS that emphasizes the group
theoretical aspect of the theory. This aspect is crucial for the following subsection \ref{sec:GenLift} where we show how to lift
Floquet solutions of the TLS to higher spins $s>1/2$. Also this lift procedure has been used before but we present
a r\'{e}sum\'{e} for the convenience of the reader. The next subsection \ref{sec:GenSO3} is vital for the remainder of the paper
insofar as it reduces the Floquet problem for the TLS to its classical limit. More precisely, the classical equation of motion
for a spin ${\mathbf X}$ in a periodic magnetic field has, in the generic case, exactly two periodic solutions $\pm{\mathbf X}(t)$,
and the Floquet solutions $u_\pm(t)$ together with the quasienergies $\epsilon_\pm$ can be derived from $\pm{\mathbf X}(t)$. This is
the content of  Assertion \ref{ass1}. The next section \ref{sec:Geo} closer investigates some geometrical aspects of
the problem. The Bloch sphere can either be viewed as the set of one-dimensional projections of the TLS or as the phase space of its
classical limit. The first view leads to a scenario that has been analyzed in \cite{AA87} and \cite{MB14} in the context of generalized Berry phases.
Following this approach in subsection \ref{sec:GeoI} we are lead to the splitting of the quasienergy into a geometrical and a dynamical part.
The classical mechanics approach in subsection \ref{sec:GeoII} shows that the quasienergy is essentially the integral
of the Poincar\'{e}-Cartan form over one closed orbit. This result is closely connected to the approach to semi-classical Floquet theory
in \cite{BH91}.

The second main part on resonances essentially bears on the resonance condition due to J.~H.~Shirley  \cite{S65}.
After a short subsection \ref{sec:H} on the quasienergy as a homogeneous function we show in subsection \ref{sec:Grad}
that the resonance condition is equivalent to the vanishing of the time-average of the $3$rd component of the classical
periodic solution ${\mathbf X}(t)$ (Assertion \ref{ass2}) and that the slope of the function $\epsilon(\omega)$,
where $\omega$ denotes the frequency of the periodic magnetic field, is equal to the geometric part of the
quasienergy divided by $\omega$ (Assertion \ref{ass3}).

The third main part deals with analytical approximations to the RPL. If the classical periodic solution ${\mathbf X}(t)$
is expanded into a Fourier series, the equation of motion can be rewritten as an infinite-dimensional matrix problem. This is
similar to the approach in \cite{AT55} and \cite{S65} to the TLS Schr\"odinger equation. Since the involved matrix $A$ and any truncation
$A^{(N)}$ of it is tridiagonal the determinant of $A^{(N)}$ and all relevant minors can be determined by recursion relations.
Thus we obtain, in subsection \ref{sec:AATF},  analytical results for the truncated Fourier series of ${\mathbf X}(t)$
that are arbitrary close to the exact solution. By means of Assertion \ref{ass1} these analytical approximations can also be
used to calculate the quasienergy in subsection \ref{sec:AAQE}.
As expected, we observe different branches and avoided level crossing at the resonance frequencies.
The latter can be approximately determined, using Assertion \ref{ass2}, via $\det A^{(N)}=0$, see subsection \ref{sec:AARF}.

The remainder of the paper, section \ref{sec:LC},
is devoted to the investigation of various limit cases that often require additional ideas for
asymptotic solutions and not simply the evaluation of the truncated Fourier series. The RPL has three parameters,
namely the Larmor frequency $\omega_0$ of the constant magnetic field into $z$-direction,
the amplitude $F$ of the periodic field into $x$-direction
and the frequency $\omega$ of the periodic field. Accordingly, there are the three limit cases where $F\rightarrow 0$,
see subsection \ref{sec:LCF},  $\omega_0\rightarrow 0$, see subsection \ref{sec:LCOM0},
and $\omega\rightarrow 0$, see subsection \ref{sec:LCOM}. Moreover, there are complementary limit cases where $\omega\rightarrow\infty$,
see subsection \ref{sec:FQW}, and $\omega_0\rightarrow\infty$, see subsection \ref{sec:FQX}.
The case $F\rightarrow \infty$ is somewhat intricate and will be treated in section \ref{sec:LCOM} and not in an own subsection.
We want to highlight three features among the various limit cases. First, by using the resonance condition in the form $\det A^{(N)}=0$
it is a straightforward task to calculate a finite number of terms of the $F$-power series for the Bloch-Siegert shifts
that can be compared with known results from the literature. Second, for small $F$ it is sensible to expand the Fourier coefficients
of ${\mathbf X}(t)$ into power series in $F$. This leads to so-called Fourier-Taylor series that are defined in-depth in subsection
\ref{sec:FT} and also give rise to analytical approximations of the quasienergy within their convergence domain. Finally, the classical RPL equation of motion has an exact ``pendulum" solution for $\omega_0=0$ that can be extended to a solution valid even in linear order w.~r.~t.~$\omega_0$.
In this order it is also possible to obtain a simple expression for the quasienergy and to solve the Schr\"odinger equation, see subsection \ref{sec:SE}.
Hence this limit case seems to be suited for further studies.

We close with a summary and outlook in section \ref{sec:SO}.

%%%%%%%%%%%%%%%%%%%%%%%%%%%%%%%%%%%%%%%%%%%%%%%%%%%%%%%%%%%%%%%%%%%%%%%%%%%%%%%%%%%%%%%%%%%%%%%%%%%%%%%%%%%%%%%%%%%%%%%%%%%%%%%
\section{Generalities}\label{sec:Gen}
%%%%%%%%%%%%%%%%%%%%%%%%%%%%%%%%%%%%%%%%%%%%%%%%%%%%%%%%%%%%%%%%%%%%%%%%%%%%%%%%%%%%%%%%%%%%%%%%%%%%%%%%%%%%%%%%%%%%%%%%%%%%%%%

%%%%%%%%%%%%%%%%%%%%%%%%%%%%%%%%%%%%%%%%%%%%%%%%%%%%%%%%%%%%%%%%%%%%%%%%%%%%%%%%%%%%%%%%%%%%%%%%%%%%%%%%%%%%%%%%%%%%%%%%%%%%%%%
\subsection{Floquet theory for $\mbox{SU}_2$}\label{sec:GenSU2}
%%%%%%%%%%%%%%%%%%%%%%%%%%%%%%%%%%%%%%%%%%%%%%%%%%%%%%%%%%%%%%%%%%%%%%%%%%%%%%%%%%%%%%%%%%%%%%%%%%%%%%%%%%%%%%%%%%%%%%%%%%%%%%%

It appears that the simplest way to explain the general ideas of Floquet theory for two level systems is by again proving its
central claim. In doing so we will emphasize the group-theoretical aspects of Floquet theory
and else will stick closely to \cite{H16}.

The Schr\"odinger equation for this system is of the form
\begin{equation}\label{Gen0}
 {\sf i}\,\frac{\partial }{\partial t}\,\psi(t)=\hat{H}(t)\,\psi(t)
 \;.
\end{equation}
Here we have set $\hbar=1$ and will assume  the Hamiltonian $\hat{H}(t)$ to be $T$-periodic in time,
\begin{equation}\label{Gen3}
 \hat{H}(t+T)=\hat{H}(t)
  \;,
\end{equation}
where throughout this paper $T=\frac{2\pi}{\omega}>0$.
(\ref{Gen0}) gives rise to a matrix equation
for the evolution operator $U(t,t_0)$ that reads
\begin{equation}\label{Gen1}
  {\sf i}\,\frac{\partial }{\partial t}\,U(t,t_0)=\hat{H}(t)\,U(t,t_0)
  \;,
\end{equation}
with initial condition
\begin{equation}\label{Gen2}
U(t_0,t_0)={\mathbbm 1}\;.
\end{equation}
We will assume that $U(t,t_0)\in SU_2$, the Lie group of unitary $2\times 2$-matrices with unit determinant.
Consequently, the Hamiltonian $\hat{H}(t)$ has to be chosen such that ${\sf i}\,\hat{H}(t)$ lies
in the corresponding Lie algebra $su_2$ of anti-Hermitean  $2\times 2$-matrices with
vanishing trace, closed under commutation $[\;,\;]$. The relation between (\ref{Gen0}) and (\ref{Gen1}) is obvious: If $\psi_0\in{\mathbbm C}^2$ and
$U(t,t_0)$ is the unique solution of (\ref{Gen1}) with initial condition (\ref{Gen2}),  then
$\psi(t)\equiv U(t,t_0)\,\psi_0$ will be the unique solution of (\ref{Gen0}) with initial condition
$\psi(t_0)=\psi_0$. Conversely, let $\psi_1(t)$ and $\psi_2(t)$ be the two solutions of (\ref{Gen0})
with initial conditions $\psi_1(t_0)={1 \choose 0}$ and $\psi_2(t_0)={0 \choose 1}$, then
$U(t,t_0)\equiv (\psi_1(t),\psi_2(t))$ will solve (\ref{Gen1}) and (\ref{Gen2}).

Further, it follows that any other solution $U_1(t,t_0)$ of (\ref{Gen1})
with initial condition $U_1(t_0,t_0)=V_0$ will be of the form
\begin{equation}\label{Gen4}
 U_1(t,t_0)= U(t,t_0)\,V_0
 \;.
\end{equation}
As a special case of (\ref{Gen4}) we consider
\begin{equation}\label{Gen5}
U_2(t,t_0)\equiv U(t+T,t_0)
\;,
\end{equation}
which, due to (\ref{Gen3}), also solves (\ref{Gen1}) but has the initial condition
\begin{equation}\label{Gen6}
  U_2(t_0,t_0)=U(t_0+T,t_0)\equiv {\mathcal F}
  \;.
\end{equation}
Hence (\ref{Gen4}) implies
\begin{equation}\label{Gen7}
  U_2(t,t_0)=U(t+T,t_0)= U(t,t_0)\, {\mathcal F}
  \;.
\end{equation}
${\mathcal F}\in\mbox{SU}_2$ is called the ``monodromy matrix". It can be written as
\begin{equation}\label{Gen8}
  {\mathcal F}=e^{-{\sf i}\, T\,F},\;{\sf i}\,F\in \mbox{su}_2
  \,.
\end{equation}
Now we define
\begin{equation}\label{Gen9}
  {\mathcal P}(t,t_0)\equiv U(t,t_0)\,e^{{\sf i}(t-t_0)F}
\end{equation}
and will show that ${\mathcal P}$ is $T$-periodic in the first argument:
\begin{eqnarray}
\label{Gen10a}
  {\mathcal P}(t+T,t_0) &=& U(t+T,t_0)\,e^{{\sf i}(t+T-t_0)F} \\
  \label{Gen10b}
   &=&  U(t+T,t_0)\,e^{{\sf i}\,T\,F}\,e^{{\sf i}(t-t_0)F}\\
   \label{Gen10c}
   &\stackrel{(\ref{Gen7},\ref{Gen8})}{=}&  U(t,t_0)\,{\mathcal F}\,{\mathcal F}^{-1}\,e^{{\sf i}(t-t_0)F}\\
   \label{Gen10d}
   &=&{\mathcal P}(t,t_0)\;.
\end{eqnarray}
Summarizing, we have shown that the evolution operator $U(t,t_0)$ can be written as the product of a periodic matrix
and an exponential matrix function of time, i.~e.~,
\begin{equation}\label{Gen11}
  U(t,t_0)= {\mathcal P}(t,t_0)\,e^{-{\sf i}(t-t_0)F}
  \;,
\end{equation}
which is essentially the Floquet theorem for two level systems.
(\ref{Gen11}) is also called the ``Floquet normal form" of $U(t,t_0)$.
For an example where an explicit solution for $U(t,t_0)$ is possible for some limit case see also subsection \ref{sec:SE}.

The derivation of (\ref{Gen11}) can be
easily generalized from $\mbox{ SU}_2$ to any other finite-dimensional matrix Lie group with the property that the exponential
map from the Lie algebra to the Lie group is surjective, since this has been implicitly used in (\ref{Gen8}).

The matrix $F$ is Hermitean and hence has an eigenbasis $|n\rangle,\,n=1,2$ and real eigenvalues $\epsilon_n$ such that
\begin{equation}\label{Gen12}
F= \epsilon_1\,|1\rangle\langle 1|+\epsilon_2\,|2\rangle\langle 2|
\;.
\end{equation}
In this eigenbasis, (\ref{Gen11}) assumes the form
\begin{eqnarray}\label{Gen13a}
 U(t,t_0)\,|n\rangle &=& {\mathcal P}(t,t_0)\,e^{-{\sf i}(t-t_0)\,F}|n\rangle\\
 \label{Gen13b}
  &=& {\mathcal P}(t,t_0)\,e^{-{\sf i}(t-t_0)\epsilon_n}|n\rangle\\
 \label{Gen13c}
 & = &{\mathcal P}(t,t_0)\,|n\rangle\,e^{-{\sf i}(t-t_0)\epsilon_n}\\
 \label{Gen13d}
  &\equiv& u_n(t,t_0)\,e^{-{\sf i}(t-t_0)\epsilon_n}
  \;,
\end{eqnarray}
where the latter functions are called ``Floquet functions" or ``Floquet solutions of (\ref{Gen0})"
and the eigenvalues $\epsilon_n$  of $F$ are called ``quasienergies", see \cite{H16}.
For the two level system we have exactly two quasienergies $\pm \epsilon$ such that $\epsilon\ge 0$ since $\mbox{Tr } F=0$.
It follows that any solution $\psi(t)$ of (\ref{Gen0})
with initial condition $\psi(t_0)=a_1\, |1\rangle +a_2\,|2\rangle$ can be written in the form
\begin{equation}\label{Gen14}
 \psi(t)= U(t,t_0)\,\psi(t_0)= \sum_{n=1}^{2}a_n\, u_n(t,t_0)\,e^{-{\sf i}(t-t_0)\epsilon_n}
 \;,
\end{equation}
with time-independent coefficients $a_n$. In this respect the $u_n(t,t_0)$, resp. $\epsilon_n$,
generalize the eigenvectors, resp.~eigenvalues,  of a time-independent Hamiltonian $\hat{H}$. The latter is trivially $T$-periodic
for every $T>0$ and hence also in this case the Floquet theorem (\ref{Gen11}) must hold. Indeed it does so with
${\mathcal P}(t,t_0)={\mathbbm 1}$ and $F=\hat{H}$.

We remark that the mere analogy between Floquet solutions and eigenvectors can be given a precise meaning by considering the ``Floquet Hamiltonian"
$K$ defined on the extended Hilbert space $L^2[0,T]\otimes {\mathbbm C}^2$, see \cite{H16}, such that the quasienergies
are recovered as the eigenvalues of $K$. This was already anticipated in \cite{AT55} and \cite{S65},
but we will not go into the details since the extended Hilbert space will not be used in the present paper.

In this account of Floquet theory we have stressed the dependence of the various definitions of the choice of an arbitrary initial time $t_0$.
It hence remains to investigate the effect of changing from $t_0$ to some other initial time $t_1$. A straightforward calculation using the semi-group
property of the evolution operator
\begin{equation}\label{Gen15}
 U(t,t_0)=U(t,t_1)\,U(t_1,t_0)
\end{equation}
gives the result
\begin{eqnarray}\nonumber
 U(t,t_1)&=& {\mathcal P}(t,t_0)\, {\mathcal P}(t_1,t_0)^{-1}\\
 \label{Gen16a}
 &&\exp\left(-{\sf i}(t-t_1) {\mathcal P}(t_1,t_0)\,F\, {\mathcal P}(t_1,t_0)^{-1}\right)\\
 \label{Gen16b}
 &\equiv&{\mathcal P}(t,t_1)\,\exp\left(-{\sf i}(t-t_1)\widetilde{F}\right)
 \;.
\end{eqnarray}
It follows that the eigenvalues of $\widetilde{F}\equiv {\mathcal P}(t_1,t_0)\,F\, {\mathcal P}(t_1,t_0)^{-1}$
and $F$ coincide, hence the change of the initial time will modify the Floquet functions but not the quasienergies.
In concrete applications there will often be a natural choice for $t_0$ and the dependence on $t_0$ may be suppressed without
danger of confusion.

We will add a few remarks on the uniqueness of the quasienergies $\epsilon_n$. It is often argued that the quasienergies are
only unique up to integer multiples of $\omega$, see, e.~g.~, \cite{H16}. It seems at first glance that in our approach
uniqueness is guaranteed by the requirement ${\sf i}\,F\in \mbox{su}_2$. For example, the replacement $\epsilon_n\mapsto \epsilon_n+\omega$
in (\ref{Gen12}) would result in $F\mapsto F+\omega\,{\mathbbm 1}$ and violate the condition $\mbox{Tr} F=0$.
But this uniqueness is achieved by using a complex $\mbox{arg}$-function with a discontinuous cut. Consider, for example, a
smooth $1$-parameter family of monodromy matrices ${\mathcal F}(\omega)$ and the corresponding family $\epsilon_1(\omega)$ of quasienergies.
It may happen that $\exp\left( -{\sf i}\,T\,\epsilon_1(\omega)\right)$ crosses the cut and hence $\epsilon_1(\omega)$ changes discontinuously.
But this discontinuity is not a physical effect and only due to the choice of the $\mbox{arg}$-function. It could be compensated by, say, passing from $\epsilon_1(\omega)$ to $\epsilon_1(\omega)+\omega$. In this case it would be more appropriate to consider, say, $\epsilon_1(\omega)$ and $\epsilon_1(\omega)+\omega$ as physically equivalent quasienergies. Generally speaking, the issue of continuity is an argument in favour
of considering the quasienergies modulo $\omega$.

%%%%%%%%%%%%%%%%%%%%%%%%%%%%%%%%%%%%%%%%%%%%%%%%%%%%%%%%%%%%%%%%%%%%%%%%%%%%%%%%%%%%%%%%%%%%%%%%%%%%%%%%%%%%%%%%%%%%%%%%%%%%%%%
\subsection{Lift to higher spins}\label{sec:GenLift}
%%%%%%%%%%%%%%%%%%%%%%%%%%%%%%%%%%%%%%%%%%%%%%%%%%%%%%%%%%%%%%%%%%%%%%%%%%%%%%%%%%%%%%%%%%%%%%%%%%%%%%%%%%%%%%%%%%%%%%%%%%%%%%%

A possible physical realization of the two level system with Hilbert space ${\mathbbm C}^2$ is a single spin with
spin quantum number $s=\frac{1}{2}$. Even if the two level system is realized in a different way it will be convenient to adopt
the language of spin systems. For example, the Lie algebra $\mbox{su}_2$ is spanned by the Pauli matrices $\sigma_i,\;i=x,y,z$
(times ${\sf i}$) or, equivalently, by the three spin operators $\hat{s}_i\equiv\frac{1}{2}\sigma_i,\;i=1,2,3$ (times ${\sf i}$).
Consequently, the Hamiltonian
$\hat{H}(t)$ can always be construed as a Zeeman term with a time-dependent dimensionless magnetic field ${\mathbf h}(t)$, namely,
\begin{equation}\label{L7}
 \hat{H}(t)={\mathbf h}(t)\cdot \hat{\mathbf s}\equiv  \sum_{i=1}^{3}h_i(t) \,\hat{s}_i
 \;,
\end{equation}
where, following the usual convention, we have omitted a minus sign.
We will outline the procedure of lifting a solution of the Schr\"odinger equation for a spin with $s=\frac{1}{2}$ in a time-dependent
magnetic field to a solution of the corresponding  Schr\"odinger equation for general $s$. For this lift the $T$-periodicity of
${\mathbf h}(t)$ is not necessary, but it will later be used to draw conclusions about the Floquet states of the system with general $s$.
The lift procedure is less known but has
already in 1987 been applied to the problem of an $N$-level system in a periodic laser field \cite{H87}. See also \cite{PS04} for
a more recent application of the lift procedure to the problem of Landau-Zener transitions in a noisy environment.
For the Bloch-Siegert shift in $s=1$ systems see also \cite{HS77}.

Let, as in section \ref{sec:GenSU2}, $t\mapsto U(t,t_0)$ be a smooth curve in $\mbox{SU}_2$ such that $U(t_0,t_0)={\mathbbm 1}$.
It follows that
\begin{equation}\label{L1}
  \left(\frac{\partial}{\partial t}U(t,t_0)\right)\; U(t,t_0)^{-1}\equiv - {\sf i}\,\hat{H}(t)\in \mbox{su}_2
  \;.
\end{equation}
 Hence the columns $\psi_1(t),\;\psi_2(t)$ of $U(t,t_0)$ are linearly independent solutions of the
Schr\"odinger equation (\ref{Gen0}).
Next we consider the well-known irreducible Lie algebra representation (shortly called ``irrep"), see, e.~g.~, \cite{H62} Chapter 10--4,
\begin{equation}\label{L3}
  r^{(s)}:\mbox{ su}_2\longrightarrow \mbox{su}_{2s+1}
\end{equation}
of $\mbox{su}_2$ parametrized by the spin quantum number $s$ such that $2s\in{\mathbbm N}$
and the corresponding irreducible group representation (also called ``irrep")
\begin{equation}\label{L4}
  R^{(s)}:\mbox{SU}_2\longrightarrow \mbox{SU}_{2s+1}\;.
\end{equation}
It follows that
\begin{equation}\label{L5}
  r^{(s)}\left({\sf i}\,\hat{s}_i\right)={\sf i}\,\hat{S}_i,\;\;i=x,y,z\;,
\end{equation}
where the $\hat{s}_i$ have been defined above as the three $s=\frac{1}{2}$ spin operators and the $\hat{S}_i$ denote
the corresponding spin operators for general $s$.

It follows from (\ref{L5}) that
\begin{equation}\label{L8}
  r^{(s)}\left( -{\sf i}\,\hat{H}(t)\right)=-{\sf i}\,{\mathbf h}(t)\cdot \hat{\mathbf S}\equiv - {\sf i}\,\sum_{i=1}^{3}h_i(t) \,\hat{S}_i
  \;.
\end{equation}
Hence
\begin{equation}\label{L9}
 U(t,t_0)^{(s)}\equiv R^{(s)}\,U(t,t_0)
\end{equation}
will be a matrix solution of the lifted time evolution equation
\begin{equation}\label{L10}
 {\sf i}\,\frac{\partial}{\partial t}\, U(t,t_0)^{(s)}={\mathbf h}(t)\cdot \hat{\mathbf S}\, U(t,t_0)^{(s)}
 \;.
\end{equation}
Note that $  U(t,t_0)^{(s)}$ is a unitary matrix and hence its columns span the general $(2s+1)$-dimensional solution space
of the lifted Schr\"odinger equation.

Next we will use that the Hamiltonian is a $T$-periodic function of time, i.~e.~, that (\ref{Gen3}) holds. Consequently,
we can apply the irrep $R^{(s)}$ to the Floquet normal form (\ref{Gen11}) of $U(t,t_0)$ and obtain
\begin{eqnarray}\label{L11a}
U(t,t_0)^{(s)}&=&R^{(s)}\left( {\mathcal P}(t,t_0)\right)\,R^{(s)}\left( e^{-{\sf i}(t-t_0)F}\right)\\
\label{L11b}
&\equiv&
{\mathcal P}(t,t_0)^{(s)}\, e^{-{\sf i}(t-t_0)F^{(s)}}
\;,
\end{eqnarray}
where
\begin{equation}\label{L12}
{\sf i}\, F^{(s)}\equiv r^{(s)} \left( {\sf i}\,F \right)
 \;.
\end{equation}
Recall that the eigenvalues of $F$ are of the form $\pm \epsilon$ where $\epsilon \ge 0$ is the quasienergy of the two level system.
Moreover, since ${\sf i}\,F\in\mbox{su}_2$, $F$ can be written in the form
\begin{equation}\label{L13}
  F=2\,\epsilon\,\sum_{i=1}^{3} f_i\,\hat{s}_i
  \;,
\end{equation}
such that $\sum_{i=1}^3 |f_i|^2=1$ and hence $\hat{s}'=\sum_{i=1}^{3} f_i\,\hat{s}_i$
is the $s=\frac{1}{2}$ spin operator into the direction $(f_1,f_2,f_3)^\top$.
From this it follows that
\begin{equation}\label{L14}
  F^{(s)}=2\,\epsilon\,\sum_{i=1}^{3} f_i\,r^{(s)}\left(\hat{s}_i\right)\stackrel{(\ref{L5})}{=}2\,\epsilon\,\sum_{i=1}^{3} f_i\, \hat{S}_i
  \;.
\end{equation}
Since $\hat{S}'=\sum_{i=1}^{3} f_i\,\hat{S}_i$
is the general $s$ spin operator into the direction $(f_1,f_2,f_3)^\top$ with eigenvalues $m=-s,\ldots,s$
it further follows that the eigenvalues of $F^{(s)}$ and hence the quasienergies of the lifted Schr\"odinger equation are of the form
\begin{equation}\label{L15}
 \epsilon_m^{(s)}=2\,\epsilon\,m,\; m=-s,\ldots,s
 \;.
\end{equation}
Also the Floquet functions for the lifted problem can be obtained from those of the two level system, and hence the
general solution of the lifted Schr\"odinger equation can be reduced to the general solution of  (\ref{Gen0}).\\

%%%%%%%%%%%%%%%%%%%%%%%%%%%%%%%%%%%%%%%%%%%%%%%%%%%%%%%%%%%%%%%%%%%%%%%%%%%%%%%%%%%%%%%%%%%%%%%%%%%%%%%%%%%%%%%%%%%%%%%%%%%%%%%
\subsection{Lift to $\mbox{SO}_3$}\label{sec:GenSO3}
%%%%%%%%%%%%%%%%%%%%%%%%%%%%%%%%%%%%%%%%%%%%%%%%%%%%%%%%%%%%%%%%%%%%%%%%%%%%%%%%%%%%%%%%%%%%%%%%%%%%%%%%%%%%%%%%%%%%%%%%%%%%%%%

We will consider the lift of the two level problem to the three level problem with spin $s=1$ in more details. To this end we will not directly use the
irrep $R^{(1)}$ but some other well-known representation $R$ that is, however, unitarily equivalent to $R^{(1)}$. It is defined by
\begin{equation}\label{SO1}
u\,\hat{s}_i\,u^\ast = \sum_{j=1}^{3}R(u)_{ij}\,\hat{s}_j,\quad \mbox{for all }i=1,2,3 \mbox{ and }u\in su_2
\;,
\end{equation}
and can be restricted to an irrep $R:\mbox{SU}_2\rightarrow \mbox{SO}_3$. The corresponding Lie algebra irrep $r:\mbox{su}_2\rightarrow \mbox{so}_3$
maps ${\sf i}\hat{s}_i,\;i=1,2,3$ onto three anti-symmetric real matrices that span  $\mbox{so}_3$.
Let
\begin{equation}\label{SO2}
  {\mathbf H}(t)\equiv
  \left(
   \begin{array}{ccc}
   0 & -h_3(t) & h_2(t) \\
  h_3(t) & 0 & -h_1(t) \\
   -h_2(t) & h_1(t) & 0
 \end{array}
   \right)\;,
\end{equation}
then the lifted evolution equation can be written as
\begin{equation}\label{SO3}
 \frac{\partial }{\partial t}\,R(t,t_0)= {\mathbf H}(t)\,R(t,t_0)
  \;,
\end{equation}
where, as usual, $R(t,t_0)\in \mbox{SO}_3$ and $R(t_0,t_0)={\mathbbm 1}$. The underlying ''Schr\"odinger equation"
has the form
\begin{equation}\label{SO4}
 \frac{d }{d t}\,{\mathbf X}(t)= {\mathbf h}(t)\times {\mathbf X}(t)
  \;,
\end{equation}
with ${\mathbf X}(t)\in{\mathbbm R}^3$ and can simultaneously be considered as the classical limit of the lifted
Schr\"odinger equation for $s\rightarrow \infty$. A more direct derivation of (\ref{SO4}) can be given in the following way.

Let $\psi(t)$ be a  solution of the Schr\"odinger equation (\ref{Gen0}) and
\begin{equation}\label{C3}
  P(t)=\left| \psi(t)\rangle\langle \psi(t)\right|
\end{equation}
be the corresponding $1$-dimensional projector.
It can also be written in the form
\begin{equation}\label{C4}
  P(t)=\frac{1}{2} {\mathbbm 1}+\sum_{i=x,y,z} X_i(t)\, \hat{s}_i
\;.
\end{equation}
We will write ${\mathbf X}(t)=(X_1(t),X_2(t),X_3(t))^\top$ and note that $0=\det P(t)=\frac{1}{4}\left(\left\| {\mathbf X}(t) \right\|^2-1\right)$.
Hence the $1$-parameter family $P(t)$ corresponds to a curve ${\mathbf X}(t)$ on the ``Bloch sphere",
defined by $\left\| {\mathbf X}(t) \right\|^2=1$.

The projector $P(t)$ satisfies the von-Neumann equation
\begin{equation}\label{C5}
\frac{d}{dt}P(t)= -{\sf i}\; \left[ \hat{H}(t),P(t)\right]\;.
\end{equation}
Using the commutation relations of the spin operators $\hat{s}_i$  one easily derives from (\ref{C5}) the differential equation
(\ref{SO4}) for ${\mathbf X}(t)$. In the special case where $\psi(t)=\psi_1(t)=u_1(t,t_0)\,e^{-{\sf i}(t-t_0)\epsilon_1}$  is a Floquet solution
of (\ref{Gen0})
the curve ${\mathbf X}(t)$ is a $T$-periodic solution of (\ref{SO4}) and can be visualized as a closed
trajectory on the Bloch sphere. The second Floquet solution $\psi_2(t)=u_2(t,t_0)\,e^{-{\sf i}(t-t_0)\epsilon_2}$
is orthogonal to $\psi_2(t)$ for all $t$ and hence must correspond to the ``antipode" periodic solution $-{\mathbf X}(t)$
of (\ref{SO4}).

It will be instructive to check the consistency of our representation by directly applying the Floquet
theory to (\ref{SO3}). The corresponding monodromy matrix ${\mathcal F}^{(1)}= R(t_0+T,t_0)=\exp\left(-{\sf i}\,T\,F^{(1)} \right)$
has, as any matrix in $\mbox{SO}_3$, three eigenvalues of the form $1,\,e^{{\sf i}\rho},\,e^{-{\sf i}\rho}$.
Consequently, $F^{(1)}$ has the eigenvalues $0, \, +\rho/T, \, -\rho/T$ which is in accordance with
(\ref{L15}) iff $2\epsilon=\rho/T$. We note in passing that these considerations suggest a simple numerical
procedure to determine the quasienergy $\epsilon$: Solve (\ref{SO3}) numerically over one period $T$ and find the eigenvalues
of the corresponding monodromy matrix  ${\mathcal F}^{(1)}$.

As a by-product of the Floquet theory for $\mbox{SO}_3$ we will prove the existence of periodic classical solutions of (\ref{SO4}).
Let ${\mathbf X}_0$ be the eigenvector of the monodromy matrix ${\mathcal F}^{(1)}= R(t_0+T,t_0)$ corresponding to the eigenvalue $1$.
If we use ${\mathbf X}_0$ as the initial value ${\mathbf X}(t_0)$ of a solution of (\ref{SO4})  we conclude
\begin{eqnarray}
\label{C6a}
  {\mathbf X}(t+T) &=& R(t+T,t_0){\mathbf X}_0 \\
   &\stackrel{(\ref{Gen11})}{=}& {\mathcal P}^{(1)}(t+T,t_0)\,e^{-{\sf i}(t+T-t_0)F^{(1)}}\,{\mathbf X}_0 \\
   \label{C6b}
   &\stackrel{(\ref{Gen8})}{=}&  {\mathcal P}^{(1)}(t,t_0)e^{-{\sf i}(t-t_0)F^{(1)}}{\mathcal F}^{(1)}{\mathbf X}_0 \\
   \label{C6c}
   &=&{\mathcal P}^{(1)}(t,t_0)\,e^{-{\sf i}(t-t_0)F^{(1)}}\,{\mathbf X}_0\\
   \label{C6d}
   &\stackrel{(\ref{Gen11})}{=}&  R(t,t_0)\,{\mathbf X}_0\\
   \label{C6e}
   &=& {\mathbf X}(t)
   \;.
\end{eqnarray}

 This means that this special solution ${\mathbf X}(t)$ will be $T$-periodic.
 We may hence ask whether it  can be obtained as the lift of a Floquet solution
 of the Schr\"odinger equation, and, if so, how this Floquet solution can be reconstructed from ${\mathbf X}(t)$.

To this end we will start with a given  $T$-periodic solution ${\mathbf X}(t)$ of (\ref{SO4}) and want to construct
a corresponding Floquet solution of (\ref{Gen0}). It is not necessary to assume the condition $\left\| {\mathbf X}(t) \right\|^2=1$ from the outset.
We may rather use the fact that (\ref{SO4}) admits the constant of motion
\begin{equation}\label{C7}
 R^2=X_1(t)^2+X_2(t)^2+X_3(t)^2\;,
\end{equation}
and hence the solutions of (\ref{SO4}) are trajectories on the Bloch sphere of radius $R$.
Then the $T$-periodic $1$-parameter family $P(t)$ of one-dimensional projectors
defined by
\begin{equation}\label{C8}
P(t)\equiv\frac{1}{2\,R}\left( \begin{array}{cc}
                            R+X_3(t) & X_1(t)-{\sf i}\, X_2(t)\\
                             X_1(t)+{\sf i}\, X_2(t) &  R-X_3(t)
                          \end{array}
                   \right)\;,
\end{equation}
satisfies the von-Neumann equation (\ref{C5}) that is equivalent to (\ref{SO4}).
Since $P(t)$ is a projector and hence satisfies $P(t)^2=P(t)$, each non-vanishing column of $P(t)$ will be an eigenvector of $P(t)$ corresponding to the eigenvalue $1$. After normalizing we thus obtain the $T$-periodic one-parameter family of vectors
\begin{equation}\label{C9}
\varphi(t)=\frac{1}{\sqrt{2 R (R+X_3(t))}}{R+X_3(t)\choose X_1(t)\,+{\sf i}\,X_2(t)}
\;,
\end{equation}
such that $|\varphi(t)\rangle\langle \varphi(t)|=P(t)$.
We note in passing that (\ref{C9}) is  undefined at the south pole of the Bloch sphere where $X_3(t)=-R$.
We cannot expect that $\varphi(t)$ is already a solution of the Schr\"odinger equation (\ref{Gen0}) but only that
$\varphi(t)$ differs from a solution $\psi(t)$ of (\ref{Gen0}) by a time-dependent phase factor $e^{{\sf i}\,\alpha(t)}$.
After some calculations using (\ref{SO4})  we obtain\\

\begin{equation}\nonumber
 \left(\hat{H}(t)-{\sf i}\,\frac{d}{dt}\right)\varphi(t)=
\end{equation}
\begin{equation}\label{C10a}
\frac{1}{2}\left(h_3(t)+\frac{h_1(t) X_1(t)+h_2(t) X_2(t)}{R+X_3(t)}\right)\, \varphi(t)
\end{equation}\\
\begin{equation}\label{C10b}
\equiv \chi\left({\mathbf X}(t) \right)\varphi(t)=\sum_{n\in{\mathbbm Z}}a_n\,e^{{\sf i}\,n\,\omega\,t}\, \varphi(t)\;,
\end{equation}
where the infinite sum in (\ref{C10b}) represents the Fourier series of the $T$-periodic function $t\mapsto\chi\left({\mathbf X}(t) \right)$.
By integrating this Fourier series over $t$ we obtain
\begin{equation}\label{C11}
 \alpha(t)\equiv a_0\,t +\sum_{\stackrel{n\in{\mathbbm Z}}{n\neq 0}}a_n\frac{e^{{\sf i}\,n\,\omega\,t}}{{\sf i}\,n\,\omega}\;,
\end{equation}
(neglecting an additional integration constant that would only yield a constant phase factor) and further
\begin{equation}\label{C12}
 \left(H(t)-{\sf i}\,\frac{d}{dt}\right)\psi(t)=0\;,
\end{equation}
where
\begin{eqnarray}\label{C13a}
\psi(t)&\equiv& \exp\left(- {\sf i}\,\alpha(t)\right)\,\varphi(t)\\
\label{C13b}
&=&\exp\left(- {\sf i}\sum_{\stackrel{n\in{\mathbbm Z}}{n\neq 0}}a_n\frac{e^{{\sf i}\,n\,\omega\,t}}{{\sf i}\,n\,\omega}\right)\,\varphi(t)\,\exp\left(- {\sf i}a_0\,t\right)\\
\label{C13c}
&\equiv& u(t) \,\exp\left(- {\sf i}\,\epsilon\,t\right)\;.
\end{eqnarray}
According to (\ref{C13c}) and (\ref{C12}), $\psi(t)$ is indeed a Floquet solution of (\ref{Gen0}) with quasienergy $\epsilon= a_0$ modulo $\omega$
since $u(t)$ is  $T$-periodic. The quasienergy  $\epsilon$ is the time average of $\chi({\mathbf X}(t))$ denoted by an overbar:
\begin{equation}\label{C13d}
 \epsilon= a_0=\overline{\frac{1}{2}\left(h_3(t)+\frac{h_1(t) X_1(t)+h_2(t) X_2(t)}{R+X_3(t)}\right)}\;.
\end{equation}

Thus we have proven the following:
\begin{ass}\label{ass1}
  There exists a $1:1$ correspondence between $T$-periodic solutions ${\mathbf X}(t)$ of (\ref{SO4}) such that $\left\| {\mathbf X}(t) \right\|^2=1$ and
  Floquet solutions $\psi(t)=u(t)\,\exp(-{\sf i}\,\epsilon\, t)$ of (\ref{Gen0}) satisfying the following conditions:\\
 (i) If $\psi(t)$ is a Floquet solution of (\ref{Gen0}) then
 \begin{equation}\label{C14}
   \left| \psi(t)\rangle\langle \psi(t)\right|=\frac{1}{2}\left( \begin{array}{cc}
                            1+X_3(t) & X_1(t)-{\sf i}\, X_2(t)\\
                             X_1(t)+{\sf i}\, X_2(t) &  1-X_3(t)
                          \end{array}
                   \right)\;,
 \end{equation}
  (ii) If ${\mathbf X}(t)$ is a normalized $T$-periodic solution of (\ref{SO4}) then
  $\psi(t)=u(t) \,\exp\left(- {\sf i}\,\epsilon\,t\right) $ will be a Floquet solution of (\ref{Gen0})
 where
  \begin{eqnarray}
  \nonumber
    u(t) &=& \exp\left(- {\sf i}\sum_{\stackrel{n\in{\mathbbm Z}}{n\neq 0}}a_n\frac{e^{{\sf i}\,n\,\omega\,t}}{{\sf i}\,n\,\omega}\right)
    \quad\times \\
     \label{C15a}
    && \frac{1}{\sqrt{2 (1+X_3(t))}}{1+X_3(t)\choose X_1(t)\,+{\sf i}\,X_2(t)}\;,\\
    \epsilon &=& \overline{\frac{1}{2}\left(h_3(t)+\frac{h_1(t) X_1(t)+h_2(t) X_2(t)}{1+X_3(t)}\right)}\;,
   \end{eqnarray}
   and the $a_n$ in (\ref{C15a}) are the Fourier coefficients of the $T$-periodic function
   $t\mapsto\chi\left({\mathbf X}(t) \right)$ defined in (\ref{C10a}) and (\ref{C10b}).
\end{ass}

%%%%%%%%%%%%%%%%%%%%%%%%%%%%%%%%%%%%%%%%%%%%%%%%%%%%%%%%%%%%%%%%%%%%%%%%%%%%%%%%%%%%%%%%%%%%%%%%%%%%%%%%%%%%%%%%%%%%%%%%%%%%%%
\subsection
{The RPC example I}\label{sec:RPCI}
%%%%%%%%%%%%%%%%%%%%%%%%%%%%%%%%%%%%%%%%%%%%%%%%%%%%%%%%%%%%%%%%%%%%%%%%%%%%%%%%%%%%%%%%%%%%%%%%%%%%%%%%%%%%%%%%%%%%%%%%%%%%%%%%%%

We will check the results of Assertion \ref{ass1} for the exactly solvable case of the
circularly polarized Rabi problem (RPC) where
\begin{equation}\label{C17}
 {\mathbf h}(t)=\left( \begin{array}{c}
                        F\,\cos\omega\,t \\
                        F\,\sin\omega\,t \\
                        \omega_0
                       \end{array}
 \right)\;.
\end{equation}
We obtain $T=\frac{2\pi}{\omega}$-periodic solutions ${\mathbf X}(t)$ of (\ref{SO4}) by the following argument:
Obviously,
\begin{equation}\label{C18}
 \frac{d{\mathbf h}}{dt}=\left( \begin{array}{c}
                       -\omega F\,\sin\omega\,t \\
                        \omega\,F\,\cos\omega\,t \\
                       0
                       \end{array}
                       \right)
                       =
                       \left( \begin{array}{c}
                       0\\
                       0\\
                       \omega
                       \end{array}
                         \right)
                       \times\left( \begin{array}{c}
                        F\,\cos\omega\,t \\
                        F\,\sin\omega\,t \\
                         \omega_0
                       \end{array}
                       \right)
                       \equiv
                       {\boldsymbol \omega}\times{\mathbf h}.
\end{equation}
We set ${\mathbf X}(t)={\mathbf h}(t)-{\boldsymbol\omega}$ and conclude
\begin{equation}\label{C19}
\frac{d{\mathbf X}}{dt}=\frac{d{\mathbf h}}{dt}\stackrel{(\ref{C18})}{=}{\boldsymbol \omega}\times{\mathbf h}
={\mathbf h}\times\left(-{\boldsymbol \omega}\right) ={\mathbf h}\times\left({\mathbf h}-{\boldsymbol \omega}\right)
={\mathbf h}\times{\mathbf X}\;,
\end{equation}
and analogously for ${\mathbf X}(t)={\boldsymbol\omega}-{\mathbf h}(t)$.
Hence one finds two $T$-periodic solution of (\ref{SO4}) of the form
\begin{equation}\label{C20}
 {\mathbf X}_\pm(t)=\pm\left( \begin{array}{c}
                        F\,\cos\omega\,t \\
                        F\,\sin\omega\,t \\
                         \omega_0-\omega
                       \end{array}
 \right)\;.
\end{equation}
In this case the function $\chi({\mathbf X}(t))$, see (\ref{C10b}),  turns out to be time-independent which directly yields the
quasienergies
\begin{equation}\label{C21}
 \epsilon_\pm=\frac{1}{2}\left( \omega\pm \Omega\right)\;,
\end{equation}
where $\Omega$ is the Rabi frequency
\begin{equation}\label{C22}
\Omega\equiv R=\sqrt{F^2+( \omega_0-\omega)^2}\;.
\end{equation}
 Moreover, the corresponding two Floquet solutions
$\psi_\pm(t)$ of the Schr\"odinger equation (\ref{Gen0}) can be obtained by (\ref{C13b})  with the result:
\begin{eqnarray}
\label{C23a}
  \psi_+(t) &=&\left(
\begin{array}{c}
 \frac{\sqrt{-\omega +\Omega +\omega_0}\, e^{-\frac{1}{2} {\sf i} t (\omega +\Omega )}}{\sqrt{2}
   \sqrt{\Omega }} \\
 \frac{F e^{\frac{1}{2}  {\sf i} t (\omega -\Omega )}}{\sqrt{2} \sqrt{\Omega  (-\omega +\Omega
   +\omega_0)}} \\
\end{array}
\right)\;, \\
\label{C23b}
  \psi_-(t) &=& \left(
\begin{array}{c}
 \frac{\sqrt{\omega +\Omega -\omega_0}\, e^{-\frac{1}{2}  {\sf i} t (\omega -\Omega )}}{\sqrt{2}
   \sqrt{\Omega }} \\
 -\frac{F e^{\frac{1}{2} {\sf i} t (\omega +\Omega )}}{\sqrt{2} \sqrt{\Omega  (\omega +\Omega
   -\omega_0)}} \\
\end{array}
\right)\;,
\end{eqnarray}
in accordance with the well-known result, see, e.~g.~, \cite{LH14}.

%%%%%%%%%%%%%%%%%%%%%%%%%%%%%%%%%%%%%%%%%%%%%%%%%%%%%%%%%%%%%%%%%%%%%%%%%%%%%%%%%%%%%%%%%%%%%%%%%%%%%%%%%%%%%%%%%%%%%%%%%%%%%%%
\section{Geometry of the two level system}\label{sec:Geo}
%%%%%%%%%%%%%%%%%%%%%%%%%%%%%%%%%%%%%%%%%%%%%%%%%%%%%%%%%%%%%%%%%%%%%%%%%%%%%%%%%%%%%%%%%%%%%%%%%%%%%%%%%%%%%%%%%%%%%%%%%%%%%%%
 The correspondence between $T$-periodic solutions ${\mathbf X}(t)$ of (\ref{SO4}) and
 Floquet solutions $\psi(t)$ of (\ref{Gen0}) that has been formulated in Assertion \ref{ass1} can be
 further analyzed w.~r.~t.~two different geometric perspectives: Either the map $\psi(t)\mapsto {\mathbf X}(t)$
 can be viewed as the restriction of the map of a Hilbert space ${\mathcal H}$
 onto the corresponding projective Hilbert space  $P({\mathcal H})$ and the quasienergy as
 a phase change during a cyclic quantum evolution in the sense of \cite{AA87}. Or the Bloch sphere can be construed as the phase
 space of the classical limit of the two level system and the quasienergy can be related to its semi-classical limit in the sense
 of \cite{BH91}. We will treat both aspects in the following subsections.

%%%%%%%%%%%%%%%%%%%%%%%%%%%%%%%%%%%%%%%%%%%%%%%%%%%%%%%%%%%%%%%%%%%%%%%%%%%%%%%%%%%%%%%%%%%%%%%%%%%%%%%%%%%%%%%%%%%%%%%%%%%%%%%
\subsection{Geometry of the fibre bundle $\pi:{\mathbbm C}^2_{\left | 1\right |}\longrightarrow {\mathcal S}^2$}\label{sec:GeoI}
%%%%%%%%%%%%%%%%%%%%%%%%%%%%%%%%%%%%%%%%%%%%%%%%%%%%%%%%%%%%%%%%%%%%%%%%%%%%%%%%%%%%%%%%%%%%%%%%%%%%%%%%%%%%%%%%%%%%%%%%%%%%%%%

The map of wave functions to projectors $\psi \mapsto \left|\psi\rangle\langle \psi\right|$ can be viewed as a surjective map $\pi$
from the unit sphere ${\mathbbm C}^2_{\left| 1 \right|}$ of ${\mathbbm C}^2$ to the unit sphere of ${\mathbbm R}^3$, i.~e.~, as
$\pi:{\mathbbm C}^2_{\left | 1\right |}\longrightarrow {\mathcal S}^2$.
Since ${\mathbbm C}^2_{\left| 1 \right|}\cong {\mathcal S}^3$ this is essentially the Hopf fibration \cite{H31}.
The fiber $\pi^{-1}({\mathbf X})$ consists of the set of all phase factors
$e^{{\sf i}\alpha}$ and can hence be identified with the $1$-dimensional unitary group $U_1$.
Then the map $\sigma:{\mathbf X}\mapsto \varphi(t)$ according to (\ref{C9}) can be viewed
as a local section of the principal fiber bundle $\pi:{\mathbbm C}^2_{\left | 1\right |}\longrightarrow {\mathcal S}^2$ with structure group $U_1$.
The above remark that $\sigma$ is undefined at the south pole of the Bloch sphere means that $\sigma$ cannot be extended to a global section due to topological obstacles.
The group $\mbox{SU}_2$ operates in a natural fashion on ${\mathbbm C}^2_{\left| 1 \right|}$ as well as on ${\mathcal S}^2$
via rotations $R(u)$ defined in (\ref{SO1})

Following \cite{AA87} we may split the function
\begin{equation}\label{C24}
\chi\left({\mathbf X}(t) \right)=\langle \varphi(t)\left| \left(\hat{H}(t)-{\sf i}\,\frac{d}{dt}\right)    \right| \varphi(t)\rangle\;,
\end{equation}
see (\ref{C10b}),
into a ``dynamical" and a "geometrical part". The dynamical part is defined as
\begin{eqnarray}\label{C25a}
\chi_d\left({\mathbf X}(t) \right)&\equiv&\langle \varphi(t)\left|\hat{H}(t)\right| \varphi(t)\rangle\\
\label{C25b}
&=& \frac{1}{2R}\left( h_1 X + h_2 Y+ h_3 Z \right)
\end{eqnarray}
and represents the expectation value of the energy. Its time average yields the dynamical part of the quasienergy:
\begin{equation}\label{C25c}
 \epsilon_d\equiv \overline{\chi_d\left({\mathbf X}(t) \right)}= \frac{1}{2R}\overline{\left( h_1 X + h_2 Y+ h_3 Z \right)}
 \;.
\end{equation}

The geometrical part of $\chi\left({\mathbf X}(t) \right)$ is defined as
\begin{eqnarray}\label{C26a}
\chi_g\left({\mathbf X}(t)\right)&\equiv&\left\langle \varphi(t)\left|-{\sf i}\frac{d}{dt} \right. \varphi(t)\right\rangle\\
\label{C26b}
&=& \frac{X(t) \dot{Y}(t)-Y(t) \dot{X}(t)}{2 R (R+Z(t))}\;.
\end{eqnarray}
Using spherical coordinates $\theta,\phi$ for the Bloch sphere with radius $1$ we may write the differential $\chi_g\,dt$ in the form
\begin{eqnarray}
\label{C27a}
\chi_g\left({\mathbf X}(t)\right)\,dt &=&  \frac{X \,dY-Y \,dX}{2 R (R+Z)}\\
 \label{C27b}
   &=& \frac{1-\cos\theta}{2}\,d\phi \equiv \alpha\;.
\end{eqnarray}
This yields a differential $1$-form $\alpha$ on the Bloch sphere and the time average of  $\chi_g\left({\mathbf X}(t)\right)$
is, up to the factor $\frac{1}{T}=\frac{\omega}{2\pi}$, the integral of $\alpha$ over the closed curve  ${\mathcal C}$ parametrized by ${\mathbf X}(t)$.
By applying Stoke's theorem we obtain
\begin{eqnarray}\label{C28a}
\epsilon_g= \overline{\chi_g\left({\mathbf X}(t)\right)}&=&\frac{\omega}{2\pi}\oint_{\mathcal C}\alpha\\
 \label{C28b}
 &=&\frac{\omega}{2\pi}\int_{\mathcal A} d\alpha  =\frac{\omega}{4\pi}\left| {\mathcal A}\right|\;,
\end{eqnarray}
where ${\mathcal A}$ denotes the oriented area enclosed by ${\mathcal C}$
and $\left| {\mathcal A}\right|$  the correspondingly signed solid angle.
Here we have used that the $2$-form
\begin{equation}\label{C29}
  d\alpha = \frac{1}{2}\sin\theta\, d\theta \wedge d\phi
\end{equation}
equals $\frac{1}{2}$ times the surface element on the unit sphere.

Thus we obtain the following interpretation of the quasienergy $\epsilon=\epsilon_d+\epsilon_g$ as composed of two parts:
The dynamical part $\epsilon_d$ is the time average of the energy (\ref{C25b})
and the geometrical part $\epsilon_g$ is $\frac{\omega}{4\pi}$  times the solid angle enclosed by the
$T$-periodic solution of (\ref{SO4}).

We refer to \cite{AA87} for the differential-geometric background of this scenario. Obviously, the $1$-form
$\alpha$ is the retract of the canonical connection $1$-form $\gamma$ of the principal fiber bundle
$\pi:{\mathbbm C}^2_{\left | 1\right |}\longrightarrow {\mathcal S}^2$ w.~r.~t.~the local section
$\sigma:{\mathcal S}^2\longrightarrow {\mathbbm C}^2_{\left | 1\right |}$ described above.
Hence the non-vanishing of $\epsilon_g$ is due to the curvature of this principal fiber bundle and generalizes Berry's phase.
Recall further that, by definition,
$\gamma$ has values in the Lie algebra of $U_1$. This Lie algebra is isomorphic to ${\mathbb R}$ and an analogous identification
has been made by considering above $\alpha$ as a usual real-valued $1$-form.
However, $\epsilon_g$ as the integral over $\alpha$  should properly have values in
$U_1$ and not in ${\mathbbm R}$ and this subtle difference is in turn in accordance with the general claim that quasienergies are only
defined modulo $\omega$.

%%%%%%%%%%%%%%%%%%%%%%%%%%%%%%%%%%%%%%%%%%%%%%%%%%%%%%%%%%%%%%%%%%%%%%%%%%%%%%%%%%%%%%%%%%%%%%%%%%%%%%%%%%%%%%%%%%%%%%%%%%%%%%
\subsection{The RPC example II}\label{sec:RPCII}
%%%%%%%%%%%%%%%%%%%%%%%%%%%%%%%%%%%%%%%%%%%%%%%%%%%%%%%%%%%%%%%%%%%%%%%%%%%%%%%%%%%%%%%%%%%%%%%%%%%%%%%%%%%%%%%%%%%%%%%%%%%%%%%%%%

We will illustrate the results of the preceding subsection by the explicitly solvable case of RPC. The calculation is essentially identical with that in \cite{AA87}.
It follows from (\ref{C17}) and (\ref{C20}) that the dynamical part of the quasienergy assumes the value (note that the involved functions
are constant and taking the time average is superfluous)
\begin{equation}\label{C29}
  \epsilon_d=\frac{h_1 X+ h_2 Y +h_3 Z}{2R}= \frac{F^2-\omega + \omega_0}{2 \sqrt{F^2+(\omega_0 - \omega)^2}}\;.
\end{equation}
On the other hand, the vector ${\mathbf X}_+(t)$ prescribes a circle on the Bloch sphere
with constant $Z= \omega_0-\omega$, see (\ref{C20}). Consider first the case of $Z>0$.
The corresponding spherical segment has an area of $2\pi R (R-Z)$ corresponding to a solid angle of
$\left| {\mathcal A}\right|=\frac{2\pi (R-Z)}{R}$, where $R=\Omega=\sqrt{F^2+(\omega_0 - \omega)^2}$ is the radius of the Bloch sphere.
Hence
\begin{equation}\label{C30}
  \epsilon_g=\frac{\omega}{4\pi}\,\frac{2\pi (R-Z)}{R}=\frac{\omega\,(R- \omega_0+\omega)}{2\,R}\;,
\end{equation}
and
\begin{eqnarray}\label{C31a}
 \epsilon_+&=&\epsilon_d+ \epsilon_g= \frac{F^2-\omega + \omega_0}{2\, R}+\frac{\omega\,(R- \omega_0+\omega)}{2\,R}\\
 \label{C31b}
&=& \frac{1}{2}\left(\omega+R\right)\;,
\end{eqnarray}
in accordance with (\ref{C21}). For the second periodic solution ${\mathbf X}_-(t)$ it follows that both terms
$\epsilon_d$ and $\epsilon_g$ acquire a minus sign, the latter since the
spherical segment encircled by ${\mathbf X}_-(t)$
has a negative orientation. Hence $ \epsilon_-=- \epsilon_+\mod \omega$ in accordance with (\ref{C21}).

In the case of $Z<0$ the dynamical part $\epsilon_d$ remains unchanged whereas the solid angle of the spherical segment
encircled by ${\mathbf X}_+(t)$ assumes the form $\left| {\mathcal A}\right|=-\frac{2\pi (R+Z)}{R}$. This yields
$\epsilon_+=\frac{1}{2}\left(-\omega+R\right)$ which differs from (\ref{C31b}) only by $-\omega$ and is thus equivalent.
Analogous arguments hold for $\epsilon_-$.

%%%%%%%%%%%%%%%%%%%%%%%%%%%%%%%%%%%%%%%%%%%%%%%%%%%%%%%%%%%%%%%%%%%%%%%%%%%%%%%%%%%%%%%%%%%%%%%%%%%%%%%%%%%%%%%%%%%%%%%%%%%%%%
\subsection{Another solvable example}\label{sec:EX}
%%%%%%%%%%%%%%%%%%%%%%%%%%%%%%%%%%%%%%%%%%%%%%%%%%%%%%%%%%%%%%%%%%%%%%%%%%%%%%%%%%%%%%%%%%%%%%%%%%%%%%%%%%%%%%%%%%%%%%%%%%%%%%%%%%

\begin{figure}[ht]
  \centering
    \includegraphics[width=1.0\linewidth]{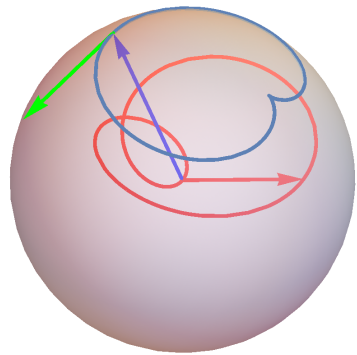}
  \caption[EX]
  {A periodic solution ${\mathbf X}(t)$ of the classical equation of motion (\ref{SO4}) according to (\ref{EX1}) with $f=1,\;\omega=1$
  represented by a closed trajectory on the Bloch sphere (blue curve).
  The corresponding magnetic field ${\mathbf h}(t)$ according to (\ref{EX2}) is visualized by the red curve. At the time $t=\pi$
  the three vectors ${\mathbf X}(\pi)$, ${\mathbf h}(\pi)$ and
  $\dot{\mathbf X}(\pi)={\mathbf h}(\pi)\times {\mathbf X}(\pi)$
  are indicated by the colored arrows.
  }
  \label{FIGEX}
\end{figure}

The idea of a ``reverse engineering of the control" ${\mathbf h}(t)$, see \cite{GDG10} -- \cite{MN14},  can also be applied to the classical Floquet problem.
Given a normalized $T$-periodic function ${\mathbf X}(t)$ we may choose ${\mathbf h}(t)\equiv {\mathbf X}(t)\times \dot{\mathbf X}(t)$
such that (\ref{SO4}) is satisfied. This special choice of ${\mathbf h}$ entails ${\mathbf h}\cdot{\mathbf X}=0$ and hence the dynamical part
$\chi_d$ of $\chi$ vanishes according to (\ref{C25b}). The geometrical part $\chi_g=\chi$ will be
calculated according to (\ref{C26b}) for the following example:
\begin{equation}\label{EX1}
{\mathbf X}(t)=\left(
\begin{array}{c}
 \cos (\omega t ) \sin \left(f \sin ^2\left(\frac{\omega t }{2}\right)\right) \\
 \sin (\omega t ) \sin \left(f \sin ^2\left(\frac{\omega t }{2}\right)\right) \\
 \cos \left(f \sin ^2\left(\frac{\omega t }{2}\right)\right) \\
\end{array}
\right)
 ,
\end{equation}
$f$ being a real parameter, which leads to
\begin{equation}\label{EX2}
{\mathbf h}(t)=\left(
\begin{array}{c}
 -\frac{1}{2} \omega  \left(f \sin ^2(\omega t )+\cos (\omega t ) \sin (f-f \cos
   (\omega t ))\right) \\
 \frac{1}{2} \omega  \sin (\omega t ) (f \cos (\omega t )-\sin (f-f \cos (t
   \omega ))) \\
 \omega  \sin ^2\left(f \sin ^2\left(\frac{\omega t }{2}\right)\right) \\
\end{array}
\right)
 ,
\end{equation}
and
\begin{equation}\label{EX3}
 \chi(t)=\omega  \sin ^2\left(\frac{1}{2} f \sin ^2\left(\frac{\omega t}{2}\right)\right)
 \;,
\end{equation}
see Figure \ref{FIGEX} visualizing the example $f=1,\;\omega=1$.
The Fourier series of $\chi(t)$ can be explicitly calculated:
\begin{equation}\label{EX4}
 \chi(t)=\epsilon+\sum_{n=1}^{\infty}b_n\,\cos n\,\omega\,t
 \;,
\end{equation}
where
\begin{eqnarray}
\label{EX5a}
  \epsilon &=& \frac{\omega}{2} \left(1-\cos \left(\frac{f}{2}\right)   J_0\left(\frac{f}{2}\right)\right)\;, \\
  \label{EX5b}
  b_n &=&\omega\,J_n\left(\frac{f}{2}\right)\,\left\{
\begin{array}{l@{\;:\;}l}
 (-1)^{\frac{n+1}{2}} \sin \left(\frac{f}{2}\right)& n\mbox{ odd },\\
 (-1)^{\frac{n+2}{2}} \cos \left(\frac{f}{2}\right) & n\mbox{ even },
\end{array}
\right. .
\end{eqnarray}
Here the $J_n(\ldots)$ denote the Bessel functions of first kind and integer order.
The constant term of (\ref{EX4}) is the quasienergy $\epsilon$ according to (\ref{EX5a}) that only depends linearly on $\omega$.
The Fourier series of $\chi(t)$
could be utilized to explicitly determine the Floquet solutions $u_\pm(t) e^{\mp{\sf i}\,\epsilon \,t}$
of the corresponding Schr\"odinger equation according to Assertion \ref{ass1}.

Note further that Assertion \ref{ass3} of section \ref{sec:Grad} can be sharpened to $\frac{\partial \epsilon}{\partial \omega}=\frac{\epsilon}{\omega}$
since $\epsilon_d=0$ in this example.

%%%%%%%%%%%%%%%%%%%%%%%%%%%%%%%%%%%%%%%%%%%%%%%%%%%%%%%%%%%%%%%%%%%%%%%%%%%%%%%%%%%%%%%%%%%%%%%%%%%%%%%%%%%%%%%%%%%%%%%%%%%%%%%
\subsection{Classical mechanics of the two level system}\label{sec:GeoII}
%%%%%%%%%%%%%%%%%%%%%%%%%%%%%%%%%%%%%%%%%%%%%%%%%%%%%%%%%%%%%%%%%%%%%%%%%%%%%%%%%%%%%%%%%%%%%%%%%%%%%%%%%%%%%%%%%%%%%%%%%%%%%%%

First we will introduce some concepts of classical mechanics suited for the present case. Let $z$ and $\varphi$ be
coordinates of the unit Bloch sphere defined by
\begin{eqnarray}
\label{PR2a}
  X/R &=& \sqrt{1-z^2}\cos\varphi\;, \\
  \label{PR2b}
  Y/R &=& \sqrt{1-z^2}\sin\varphi\;,\\
  \label{PR2c}
  Z/R &=& z\;.
\end{eqnarray}
Further we define the classical Hamiltonian
\begin{eqnarray} \label{PR3a}
  H&=&\frac{1}{R}\left(h_1\,X+h_2\,Y+h_3\,Z\right)\\
  \label{PR3b}
  &=& \sqrt{1-z^2}\left( h_1\,\cos\varphi+h_2\,\sin\varphi\right)+h_3\,z\;,
\end{eqnarray}
and rewrite the differential equation  (\ref{SO4})  in terms of the two functions $z(t)$ and $\varphi(t)$:
\begin{eqnarray}
\label{PR4a}
 \dot{z} &=& \sqrt{1-z^2}\left(h_1\,\sin\varphi- h_2\,\cos\varphi\right)=-\,\frac{\partial H}{\partial \varphi}\;, \\
\nonumber
 \dot{\varphi} &=& h_3-\frac{z}{ \sqrt{1-z^2}}\left( h_1\,\cos\varphi+h_2\,\sin\varphi\right)=\frac{\partial H}{\partial z}\;.\\
 &&\label{PR4b}
\end{eqnarray}
Note that due to (\ref{C25b}), $H$ is twice the expectation value of the Hamiltonian $\hat{H}$.
Obviously, (\ref{PR4a}) -- (\ref{PR4b}) can be viewed as Hamiltonian equations of motions in a two-dimensional
phase space isomorphic to ${\mathcal S}^2$ with canonical coordinates
\begin{equation}\label{PR5}
 p=z,\quad q=\varphi\;.
\end{equation}
We will henceforward often use $p$ and $q$ instead of $z$ and $\varphi$. Following \cite{A89} we consider
the extended phase space ${\mathcal P}={\mathcal S}^2 \times {\mathbbm R}$ with coordinates $p,\,q,\,t$
and the Poincar\'{e}-Cartan form
\begin{equation}\label{PR6}
 {\boldsymbol\alpha}\equiv p\,dq-H\,dt
 \end{equation}
defined on ${\mathcal P}$ (not to be confounded with the $1$-form $\alpha$ defined in section \ref{sec:GeoII}).
The Hamiltonian equations (\ref{PR4a}), (\ref{PR4b}) can be geometrically construed as the
direction field on ${\mathcal P}$ that is given by the unique null direction field of the exterior derivative of the Poincar\'{e}-Cartan form
\begin{equation}\label{PR7}
 {\boldsymbol\omega}\equiv d{\boldsymbol\alpha} = dp\wedge dq-dH\wedge dt\;,
 \end{equation}
see section $44$ of \cite{A89} for the details.

Periodic solutions of (\ref{PR4a}), (\ref{PR4b}) correspond to
curves $\gamma$ in ${\mathcal P}$ that are not closed since
after one period $T$ the coordinate $t$ has changed from $t=0$ to $t=T$.
This can be repaired by defining another extended phase space ${\mathcal P}'$ via identifying points with $t$-coordinates that differ by an integer multiple of $T$, or, more formally,
\begin{equation}\label{PR8}
{\mathcal P}'\equiv{\mathcal S}^2 \times {\mathbbm R}/T{\mathbbm Z}\;.
 \end{equation}
Of course, ${\mathbbm R}/T{\mathbbm Z}$ is isomorphic to ${\mathcal S}^1$. Locally the manifolds ${\mathcal P}$ and ${\mathcal P}'$
are isomorphic, but globally they are not. The differential forms ${\boldsymbol\alpha}$ and ${\boldsymbol\omega}$ can be transferred
to ${\mathcal P}'$ since all functions involved in  ${\boldsymbol\alpha}$ and ${\boldsymbol\omega}$ are $T$-periodic. Now periodic solutions of (\ref{PR4a}), (\ref{PR4b}) correspond to closed curves $\gamma$ in ${\mathcal P}'$.

Next we will rewrite the expression (\ref{C13d}) for the quasienergy $\epsilon$. By (\ref{C25b}) and (\ref{C27b}) we obtain
\begin{eqnarray}
\label{PR9a}
  \chi_d\,dt &=&\frac{1}{2} H\,dt\;, \\
  \label{PR9b}
   \chi_g\,dt &=& \frac{1}{2}\left(1-z\right)\,d\varphi=\frac{1}{2}\left(1-p\right)\,dq\;.
\end{eqnarray}
Hence
\begin{equation}\label{PR10}
 \epsilon=\overline{\chi_g+\chi_d}=- \frac{\omega}{4\pi}\oint_\gamma \left(p\,dq-H\,dt\right) +\frac{n}{2}\omega\;,
\end{equation}
where the $n$ in the last term denotes the winding number of $\gamma$ around the $z$-axis.
This result is in close analogy to equation (2.35) of \cite{BH91} that represents the semi-classical limit
of the quasienergy for integrable Floquet systems. It thus seems that for the two level system, similar as in the case of the driven harmonic oscillator,
the semi-classical limit of the quasienergy and the exact quantum-theoretical expression coincide.
However, it has not yet been shown that the quantization procedure adopted in \cite{BH91} yields the quantum two level system
when starting from its classical limit.

%%%%%%%%%%%%%%%%%%%%%%%%%%%%%%%%%%%%%%%%%%%%%%%%%%%%%%%%%%%%%%%%%%%%%%%%%%%%%%%%%%%%%%%%%%%%%%%%%%%%%%%%%%%%%%%%%%%%%%%%%%%%%%%
\section{Resonances}\label{sec:R}
%%%%%%%%%%%%%%%%%%%%%%%%%%%%%%%%%%%%%%%%%%%%%%%%%%%%%%%%%%%%%%%%%%%%%%%%%%%%%%%%%%%%%%%%%%%%%%%%%%%%%%%%%%%%%%%%%%%%%%%%%%%%%%%

In this and the following sections we will restrict the Hamiltonian (\ref{L7}) to the following special case
\begin{eqnarray}
\label{R1a}
  h_1 &=& F \cos\omega t\;, \\
  h_2 &=& G \sin\omega t\;, \\
  h_3 &=& \omega_0\;,
\end{eqnarray}
that will be referred to as the Rabi problem with elliptical polarization (RPE). It includes the two limit cases $G\rightarrow 0$, the
Rabi problem with linear polarization (RPL),  and $G\rightarrow F$, the Rabi problem with circular polarization (RPC).
Hence the quasienergy $\epsilon$ can be written as a function $\epsilon(\omega_0,F,G,\omega)$ of the four parameters
$\omega_0,F,G,\omega$ that will be assumed to have positive values.

The corresponding classical Hamiltonian (\ref{PR3b}) reads
\begin{equation}\label{R1b}
 H(z,\varphi)=\sqrt{1-z^2}\left( F\,\cos\omega t\,\cos\varphi+G\sin\omega t\,\sin\varphi\right)+\omega_0\,z
 \;.
\end{equation}

If the two level system is coupled to a second weak electromagnetic field there may occur transitions between the two different
Floquet states, analogously as in the case of two energy levels for a time-independent Hamiltonian. J.~H.~Shirley has computed
the time-averaged probability $\overline{P}$ of such a transition, see equation (26) in \cite{S65}, with the remarkably simple result
\begin{equation}\label{R1}
  \overline{P} =\frac{1}{2}\left( 1-4 \left(\frac{\partial \epsilon}{\partial \omega_0}\right)^2\right)
  \;.
\end{equation}
Although Shirley's derivation of (\ref{R1}) refers to the RPL case, see equation (1) in \cite{S65},
one can easily check that it also holds in the more general RPE case. It implies that the transition probability
assumes its maximal value $\overline{P}_{max}=\frac{1}{2}$ for
\begin{equation}\label{R2}
  \frac{\partial \epsilon}{\partial \omega_0}=0
  \;.
\end{equation}
Hence the condition (\ref{R2}) will be called the ``resonance condition". It will be further analyzed in the following subsections.

%%%%%%%%%%%%%%%%%%%%%%%%%%%%%%%%%%%%%%%%%%%%%%%%%%%%%%%%%%%%%%%%%%%%%%%%%%%%%%%%%%%%%%%%%%%%%%%%%%%%%%%%%%%%%%%%%%%%%%%%%%%%%%
\subsection{Homogeneity of the quasienergy}\label{sec:H}
%%%%%%%%%%%%%%%%%%%%%%%%%%%%%%%%%%%%%%%%%%%%%%%%%%%%%%%%%%%%%%%%%%%%%%%%%%%%%%%%%%%%%%%%%%%%%%%%%%%%%%%%%%%%%%%%%%%%%%%%%%%%%

For the RPE  the classical equation of motion (\ref{SO4}) reduces to
\begin{eqnarray}
\label{H1a}
 \dot{X}  &=& G\,\sin (\omega t)\, Z-\omega_0\,Y\;, \\
 \label{H1b}
  \dot{Y}&=& \omega_0\,X-F\,\cos (\omega t)\, Z\;,\\
  \label{H1c}
  \dot{Z} &=& F\,\cos (\omega t)\, Y- G\,\sin (\omega t)\, X\;.
\end{eqnarray}
It is invariant under the transformation
\begin{eqnarray}\label{H2a}
  \omega_0 &\mapsto& \lambda\, \omega_0\;,\\
  \label{H2b}
   F &\mapsto& \lambda \,F\;,\\
  \label{H2c}
  G &\mapsto& \lambda \,G\;,\\
  \label{H2d}
   \omega &\mapsto& \lambda\, \omega\;,\\
  \label{H2e}
  t &\mapsto& \frac{1}{\lambda}\, t\;,
 \end{eqnarray}
for all $\lambda>0$. Under this transformation the quasienergy (\ref{C13d}) scales with $\lambda$ and hence is a positively homogeneous
function of degree $1$:
\begin{equation}\label{H3}
\epsilon(\lambda\,\omega_0,\lambda\,F,\lambda\,G,\lambda\,\omega)= \lambda\,\epsilon(\omega_0,F,G,\omega)
\end{equation}
for all $\lambda>0$. This could be used to eliminate the variable $\omega_0$ (by chosing $\lambda=\omega_0^{-1}$ which transforms $\omega_0$ into $1$)
but for some purposes, e.~g., the investigation of the limit $\omega_0\rightarrow 0$
or the analysis of the resonance condition (\ref{R2}), the elimination of $\omega_0$ is not appropriate.

By the Euler theorem the positive homogeneity of $\epsilon$ implies
\begin{equation}\label{H4}
 \epsilon(\omega_0,F,G,\omega)=\omega_0\,\frac{\partial \epsilon}{\partial \omega_0}+F\,\frac{\partial \epsilon}{\partial F}
 +G\,\frac{\partial \epsilon}{\partial G}
 +\omega\,\frac{\partial \epsilon}{\partial \omega}\;.
\end{equation}

%%%%%%%%%%%%%%%%%%%%%%%%%%%%%%%%%%%%%%%%%%%%%%%%%%%%%%%%%%%%%%%%%%%%%%%%%%%%%%%%%%%%%%%%%%%%%%%%%%%%%%%%%%%%%%%%%%%%%%%%%%%%%%
\subsection{Calculation of $\nabla\,\epsilon$}\label{sec:Grad}
%%%%%%%%%%%%%%%%%%%%%%%%%%%%%%%%%%%%%%%%%%%%%%%%%%%%%%%%%%%%%%%%%%%%%%%%%%%%%%%%%%%%%%%%%%%%%%%%%%%%%%%%%%%%%%%%%%%%%%%%%%%%%

We first consider $\frac{\partial \epsilon}{\partial \omega_0}$.
For the sake of simplicity we assume that the classical equation of motion (\ref{H1a}) -- (\ref{H1c}) for the
RPE has two unique normalized $T$-periodic solutions of the form
$\pm{\mathbf X}(t)$. The latter assumption is, for example, violated in the case $\omega_0=0$, see section \ref{sec:LCOM0},
where we have a $1$-parameter family of periodic solutions and a quasienergy $\epsilon=0$. Under these assumptions we
will prove  the following:
\begin{ass}\label{ass2}
\begin{equation}\label{Grad1}
\frac{\partial \epsilon}{\partial \omega_0}=\overline{\frac{Z(t)}{2\,R}}\;.
\end{equation}
 Hence the resonance condition
 $\frac{\partial \epsilon}{\partial \omega_0}=0$ is equivalent to $\overline{Z(t)}=0$.
\end{ass}
For the proof of this assertion we will adopt the language of classical mechanics introduced in \ref{sec:GeoII}.
In order to calculate $\frac{\partial \epsilon}{\partial \omega_0}$ we have to vary the parameter $\omega_0$.
To this end we define still another extension of the phase space by
\begin{equation}\label{PR11}
 {\mathcal P}''\equiv{\mathcal S}^2 \times {\mathbbm R}/T{\mathbbm Z}\times {\mathbbm R}_{>0}\;,
\end{equation}
with coordinates $p,q,t$ and $\omega_0>0$. Again the differential forms ${\boldsymbol\alpha}$ and ${\boldsymbol\omega}$ can be transferred to
${\mathcal P}''$ but instead of $d{\boldsymbol\alpha}={\boldsymbol\omega}$ we now have
\begin{equation}\label{PR12}
  d{\boldsymbol\alpha}={\boldsymbol\omega} - \frac{\partial H}{\partial \omega_0}\,d\omega_0\wedge dt
  \stackrel{(\ref{R1b})}{=}{\boldsymbol\omega} - z\,d\omega_0\wedge dt\;.
\end{equation}

\begin{figure}[ht]
  \centering
    \includegraphics[width=1.0\linewidth]{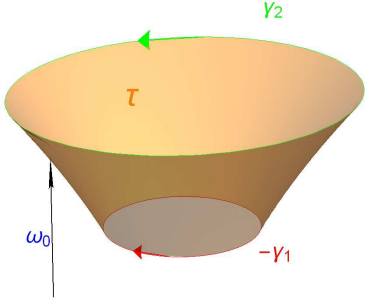}
  \caption[BS1]
  {The tube $\tau$ in extended phase space $ {\mathcal P}''$  generated by the family of closed curves $\gamma(\omega_0)$, where
  $\omega_0^{(1)}\le \omega_0 \le \omega_0^{(2)}$. The two curves $-\gamma_1$ and $\gamma_2$ that form the boundary of $\tau$ are displayed together with their orientations.
  }
  \label{FIGPR1}
\end{figure}

The closed curves $\gamma$ corresponding to periodic solutions of (\ref{PR4a}), (\ref{PR4b})
smoothly depend on $\omega_0$ and hence will be denoted by $\gamma(\omega_0)$. Geometrically, this defines a tube $\tau$ in ${\mathcal P}''$
parametrized by $\gamma(\omega_0,t)$, see Figure \ref{FIGPR1}. We will consider values of $\omega_0$ running through some closed interval
$\omega_0\in[ \omega_0^{(1)},\omega_0^{(2)}]$
and restrict the tube $\tau$ to these values. Hence the boundary $\partial \tau$ of the tube can be identified with $\gamma_2 - \gamma_1$,
where $\gamma_i \equiv \gamma(\omega_0^{(i)}),\;i=1,2$ and the minus sign in $\gamma_2 - \gamma_1$ accounts for the correct orientation.
We consider the difference between the quasienergies
\begin{equation}\label{PR13}
  \epsilon_2- \epsilon_1 \equiv \epsilon(\omega_0^{(2)})- \epsilon(\omega_0^{(1)})\stackrel{(\ref{PR10})}{=}
  - \frac{\omega}{4\pi}\left(\oint_{\gamma_2} {\boldsymbol\alpha}-\oint_{\gamma_1} {\boldsymbol\alpha}\right)+\frac{\delta n}{2}\omega\;,
\end{equation}
where $\delta n$ denotes the difference of the winding numbers. In this paper we will make the general assumption that the quasienergy can be chosen as a smooth function $\epsilon(\omega_0)$ of $\omega_0$ taking into account that it is only defined up to integer multiples of $\omega$.
Hence a discontinuous change of the winding number by an \textit{even} number can be compensated by the choice of the right branch of $\epsilon(\omega_0)$.
In the RPL example there are only even changes of the  winding number due to the symmetry of the periodic solution but,
according to our general assumption, this must hold generally.
Hence we can neglect the term $\frac{\delta n}{2}\omega$ in (\ref{PR13}).

Then
\begin{eqnarray}\label{PR14a}
  \epsilon_2- \epsilon_1&=&
  - \frac{\omega}{4\pi}\left(\oint_{\gamma_2} {\boldsymbol\alpha}-\oint_{\gamma_1} {\boldsymbol\alpha}\right)\\
  \label{PR14b}
  &=&  - \frac{\omega}{4\pi}\left(\oint_{\gamma_2-\gamma_1} {\boldsymbol\alpha}\right)\\
  \label{PR14c}
  &=& - \frac{\omega}{4\pi}\left(\oint_{\partial\tau} {\boldsymbol\alpha}\right)\\
  \label{PR14d}
  &=& - \frac{\omega}{4\pi}\left(\int_{\tau} d{\boldsymbol\alpha}\right)\\
  \label{PR14e}
  &\stackrel{(\ref{PR12})}{=}&- \frac{\omega}{4\pi}\left(\int_{\tau} {\boldsymbol\omega}-\int_{\tau} (z\,d\omega_0\wedge dt)\right)\;,
\end{eqnarray}
invoking Stokes theorem in (\ref{PR14d}).
Now we use the fact that $\int_{\tau} {\boldsymbol\omega}=0$ since the tangent plane at any point $x\in\tau$ contains a null vector
of ${\boldsymbol\omega}$, namely the vector tangent to the curve $\gamma$ passing through $x\in\tau$.
Here we have employed the above-mentioned results of analytical mechanics according to \cite{A89}.
It follows that
\begin{eqnarray}\label{PR15a}
    \epsilon_2- \epsilon_1&=& \frac{\omega}{4\pi}\int_{\tau} z\,d\omega_0\wedge dt\\
    \label{PR15b}
    &=&\frac{1}{2} \int_{\omega_0^{(1)}}^{\omega_0^{(2)}}\overline{z(\omega_0,t)}\, d\omega_0\;,
\end{eqnarray}
where $\overline{z(\omega_0,t)}$ denotes the time average w.~r.~t.~the curve $\gamma(\omega_0)$ and hence depends on $\omega_0$.
Choosing $\omega_0^{(1)}=\omega_0-\delta\,$ and $\omega_0^{(2)}=\omega_0+\delta\,$ we obtain in the limit $\delta\rightarrow 0$:
\begin{equation}\label{PR16}
 \frac{\partial \epsilon}{\partial \omega_0}= \frac{1}{2}\overline{z(\omega_0,t)}\;,
\end{equation}
which concludes the proof of Assertion \ref{ass2}.

The calculation of $\frac{\partial \epsilon}{\partial F}$ and $\frac{\partial \epsilon}{\partial G}$ can be performed
analogously. For example, the analogue of (\ref{PR15b}) reads
\begin{eqnarray}\label{PR17a}
    \epsilon_2- \epsilon_1&=& \frac{\omega}{4\pi}\int_{\tau} \frac{\partial H}{\partial F}\,dF\wedge dt\\
    \label{PR17b}
    &=&\frac{1}{2} \int_{F^{(1)}}^{F^{(2)}}\overline{\sqrt{1-z^2}\cos\omega t\,\cos\varphi}\, dF\\
    \label{PR17c}
    &=&\frac{1}{2} \int_{F^{(1)}}^{F^{(2)}}x_c\,\overline{\cos^2\omega t}\, dF\\
   \label{PR17d}
    &\rightarrow&\frac{(F^{(2)}-F^{(1)})\,x_c}{4}\;,
\end{eqnarray}
and hence
\begin{equation}\label{PR18}
  \frac{\partial \epsilon}{\partial F}=\frac{x_c}{4}\;,
\end{equation}
where $x_c$ is the coefficient of the term $\cos\omega t$ in the Fourier series of $x(t)$.
$x_c$ depends on $F$ and hence (\ref{PR17d}) only holds asymptotically for $F^{(2)}-F^{(1)}\rightarrow 0$.

Similarly,
\begin{equation}\label{PR19}
  \frac{\partial \epsilon}{\partial G}=\frac{y_s}{4}\;,
\end{equation}
where $y_s$ is the coefficient of the term $\sin\omega t$ in the Fourier series of $y(t)$.

\begin{figure}[ht]
  \centering
    \includegraphics[width=1.0\linewidth]{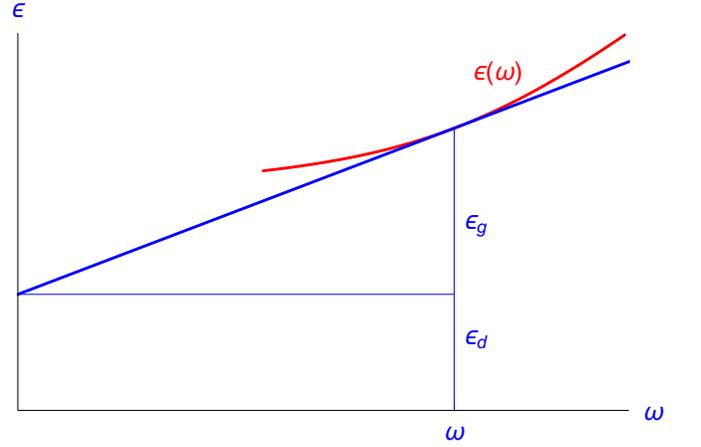}
  \caption[H1]
  {The quasienergy $\epsilon$ as a function of $\omega$ for fixed $\omega_0,\,F$ and $G$ is considered as the sum of the geometrical part
  $\epsilon_g$ and the dynamical part $\epsilon_d$ satisfying (\ref{PR23d}). Hence the values of  $\epsilon_g$ and
   $\epsilon_d$ can be geometrically determined by the intersection of the tangent to the graph of $\epsilon(\omega)$ with the $\epsilon$ axis.
  }
  \label{FIGH1}
\end{figure}

In order to calculate $\frac{\partial \epsilon}{\partial \omega}$ it is advisable first to simplify the $\omega$-dependence
of $\epsilon$ by the coordinate transformation $t\mapsto \tau \equiv \omega\,t$ together with $H\mapsto {\sf H}\equiv \frac{H}{\omega}$.
The latter transformation insures that the Hamiltonian equations of motion retain their canonical form
\begin{eqnarray}
\label{PR20a}
  \frac{d\,\varphi}{d\,\tau} &=& \frac{\partial {\sf H}}{\partial z}\;, \\
  \label{PR20b}
  \frac{d\,z}{d\,\tau} &=& -\frac{\partial {\sf H}}{\partial \varphi}\;.
\end{eqnarray}
After the coordinate transformation the Poincar\'{e}-Cartan form reads
\begin{equation}\label{PR21}
 {\boldsymbol\alpha}= p\,dq - {\sf H}\,d\tau
 \;.
\end{equation}
Recall that according to (\ref{PR10})
\begin{equation}\label{PR22}
 \epsilon(\omega)= - \frac{\omega}{4\pi}\oint_\gamma \left(p\,dq-{\sf H}\,d\tau\right) +\frac{n}{2}\omega\;.
\end{equation}
Together with an analogous calculation as in the proof of Assertion \ref{ass2} this implies
\begin{eqnarray}
\label{PR23a}
  \frac{\partial \epsilon}{\partial \omega} &=&\frac{\epsilon}{\omega} +\frac{\omega}{4\pi}\int_{0}^{2\pi}
  \frac{\partial {\sf H}}{\partial \omega}\,d\tau   \\
  \label{PR23b}
 &=& \frac{\epsilon}{\omega} -\frac{1}{4\pi\omega}\int_{0}^{2\pi} Hd\tau  \\
 \label{PR23c}
  &=& \frac{\epsilon}{\omega} -\frac{1}{2\omega}\overline{H}\\
 \label{PR23d}
  &\stackrel{(\ref{C25c})}{=}&\frac{\epsilon_g}{\omega}\;,
 \end{eqnarray}
using $\frac{\partial {\sf H}}{\partial \omega}=-\frac{1}{\omega^2}\,H$  in (\ref{PR23b}).
An alternative derivation of (\ref{PR23c}) would consist in the evaluation of the Euler relation (\ref{H4})
using (\ref{Grad1}), (\ref{PR18}), and (\ref{PR19}). Hence the calculation of the partial derivatives of $\epsilon$
does not lead to a simplified formula for $\epsilon$ itself.   However,  the relation (\ref{PR23d})
can be used for a geometrical interpretation of the splitting $\epsilon=\epsilon_g+\epsilon_d$, see Figure \ref{FIGH1}.
We will formulate this result separately and stress the fact that the above proof is independent the particular form (\ref{R1a}) of $H$.
\begin{ass}\label{ass3}
Under the assumptions of sections \ref{sec:Gen} and \ref{sec:Geo} the following holds:
\begin{equation}\label{assertion3}
 \frac{\partial \epsilon}{\partial \omega}=\frac{\epsilon_g}{\omega}\;.
\end{equation}
Hence $\frac{\partial \epsilon}{\partial \omega}$ equals the solid angle $\left|{\mathcal A}\right|$
encircled by the corresponding periodic solution of the classical RPE equation divided by $4\pi$.
\end{ass}

%%%%%%%%%%%%%%%%%%%%%%%%%%%%%%%%%%%%%%%%%%%%%%%%%%%%%%%%%%%%%%%%%%%%%%%%%%%%%%%%%%%%%%%%%%%%%%%%%%%%%%%%%%%%%%%%%%%%%%%%%%%%%%
\subsection{The RPC example III}\label{sec:RPCIII}
%%%%%%%%%%%%%%%%%%%%%%%%%%%%%%%%%%%%%%%%%%%%%%%%%%%%%%%%%%%%%%%%%%%%%%%%%%%%%%%%%%%%%%%%%%%%%%%%%%%%%%%%%%%%%%%%%%%%%%%%%%%%%%%%%%
The quasienergies
\begin{equation}\label{RPCIII0}
  \epsilon_\pm\stackrel{(\ref{C21},\ref{C22})}{=} \frac{1}{2}\left(\omega\pm \sqrt{F^2+\left( \omega_0-\omega\right)^2}\right)
\end{equation}
are obviously positively homogeneous functions of $\omega_0,F,\omega$.

Further, the normalized $3$rd component of ${\mathbf X}(t)$ has the constant value
\begin{equation}\label{RPCIII1}
  z= \frac{\omega_0-\omega}{\sqrt{F^2+\left( \omega_0-\omega\right)^2}}= \overline{z(t)}\;,
\end{equation}
see (\ref{C20}). On the other hand, by (\ref{RPCIII0}),
\begin{equation}\label{RPCIII2}
  \frac{\partial \epsilon_+}{\partial \omega_0}= \frac{1}{2} \frac{\partial \Omega}{\partial \omega_0}=
  \frac{1}{2} \frac{\omega_0-\omega}{\sqrt{F^2+\left( \omega_0-\omega\right)^2}}\;,
\end{equation}
which confirms Assertion \ref{ass2}. The resonance condition $\frac{\partial \epsilon_+}{\partial \omega_0}=0$ is equivalent
to $\omega_{res}=\omega_0$.

Moreover, Assertion \ref{ass3} is confirmed by the following calculation:
\begin{eqnarray}
\label{RPCIII2a}
  \frac{\partial \epsilon_+}{\partial \omega} &\stackrel{(\ref{RPCIII0})}{=}& \frac{1}{2} \left(1+\frac{\omega -\omega_0}
  {\sqrt{F^2+\left(\omega -\omega_0\right){}^2}}\right) \\
  \label{RPCIII2b}
   &=& \frac{1}{2} \left(1+\frac{\omega -\omega _0}{R}\right) \\
   \label{RPCIII2b}
   &\stackrel{(\ref{C30})}{=}&\frac{\epsilon_g}{\omega}
   \;.
\end{eqnarray}

%%%%%%%%%%%%%%%%%%%%%%%%%%%%%%%%%%%%%%%%%%%%%%%%%%%%%%%%%%%%%%%%%%%%%%%%%%%%%%%%%%%%%%%%%%%%%%%%%%%%%%%%%%%%%%%%%%%%%%%%%%%%%%%
\section{Analytical approximations}\label{sec:AA}
%%%%%%%%%%%%%%%%%%%%%%%%%%%%%%%%%%%%%%%%%%%%%%%%%%%%%%%%%%%%%%%%%%%%%%%%%%%%%%%%%%%%%%%%%%%%%%%%%%%%%%%%%%%%%%%%%%%%%%%%%%%%%%%

%%%%%%%%%%%%%%%%%%%%%%%%%%%%%%%%%%%%%%%%%%%%%%%%%%%%%%%%%%%%%%%%%%%%%%%%%%%%%%%%%%%%%%%%%%%%%%%%%%%%%%%%%%%%%%%%%%%%%%%%%%%%%%%
\subsection{Truncated Fourier series solution}\label{sec:AATF}
%%%%%%%%%%%%%%%%%%%%%%%%%%%%%%%%%%%%%%%%%%%%%%%%%%%%%%%%%%%%%%%%%%%%%%%%%%%%%%%%%%%%%%%%%%%%%%%%%%%%%%%%%%%%%%%%%%%%%%%%%%%%%%%

In this and the following sections we specialize to the Rabi problem with linear polarization (RPL).
Thus the classical equation of motion (\ref{SO4}) reduces to
\begin{eqnarray}
\label{AA1a}
 \dot{X}  &=&-\omega_0\,Y\;, \\
 \label{AA1b}
  \dot{Y}&=& \omega_0\,X-F\,\cos (\omega t)\, Z\;,\\
  \label{AA1c}
  \dot{Z} &=& F\,\cos (\omega t)\, Y
  \;.
\end{eqnarray}

According to our general approach we are looking for $T$-periodic solutions of these equations.
The space of real $T$-periodic functions is spanned by the four subspaces defined by even/odd $\sin$/$\cos$-series.
Assume, for example, that $X(t)$ is given by an odd $\cos$-series.
Then it follows that  $\dot{X}(t)$ is an odd $\sin$-series and, by (\ref{AA1a}), also $Y(t)$ will be an odd $\sin$-series.
By (\ref{AA1c}) we further conclude that $\dot{Z}(t)$ is an even $\sin$-series and hence $Z(t)$ will be an even $\cos$-series.
One easily derives that also both sides of (\ref{AA1b}) are odd $\cos$-series, and hence probably there exists a solution of
(\ref{AA1a}) -- (\ref{AA1c}) that fulfills the above requirements concerning the subspaces in which $X,Y$ and $Z$ lie.

On the basis of these considerations and numerical investigations we obtain the following ansatz of a (not necessarily normalized)
Fourier series solution of (\ref{AA1a}) -- (\ref{AA1c}):
\begin{eqnarray}
\label{F1a}
  X(t) &=& \sum_{n=0}^{\infty}\omega_0\, x_{2n+1}\, \cos(2n+1)\omega t\;, \\
  \label{F1b}
 Y(t) &=& \sum_{n=0}^{\infty} x_{2n+1}\,(2n+1) \,\omega\,\sin(2n+1)\omega t\;, \\
 \label{F1c}
 Z(t) &=&z_0+\sum_{n=1}^{\infty} x_{2n}\, \cos 2n\omega t\;.
\end{eqnarray}
The form of (\ref{F1b}) is already uniquely determined by the differential equation $\dot{X}=-\omega_0\,Y$.
The Fourier coefficients $x_{2n}$ of $Z(t)$ in (\ref{F1c}) are written in such a way that the vector of unknown Fourier coefficients assumes the form
 ${\mathbf x}=(x_1,x_2,\ldots)$. The validity of the ansatz (\ref{F1a}) -- (\ref{F1c}) will not be rigorously proven but appears highly
 plausible due to the investigation of the analytical approximations to these solutions in what follows.

 If we insert the ansatz (\ref{F1a}) -- (\ref{F1c})  into (\ref{AA1a}) --  (\ref{AA1c}) we
obtain an infinite system of linear equations of the form $A\,{\mathbf x}={\mathbf f}$, where
\begin{equation}\label{F2}
  {\mathbf f}=\left(\begin{array}{c}
                      -F\, z_0 \\
                      0 \\
                      \vdots
                    \end{array}
  \right)\;.
\end{equation}
The matrix $A$ is tridiagonal due to the simple form of the $h_1=F\,\cos\omega t$ which couples only neighboring modes.
Although $A$ is unbounded it may be sensible to truncate it to some $N\times N$-matrix $A^{(N)}$ if the resulting finite
Fourier series has rapidly decreasing coefficients and hence represents a good analytic approximation to the infinite Fourier series.
The matrix elements of $A$ are given by
\begin{equation}\label{AA2}
 A_{n m}=\left\{
\begin{array}{l@{\;:\;}l}
 n^2\omega^2-\omega_0^2 & n=m\mbox{ odd },\\
 -n\omega & n=m\mbox{ even },\\
\frac{F}{2} & n=m\pm 1\mbox{ odd },\\
-\frac{mF\omega}{2} & n=m\pm 1\mbox{ even },\\
 0 & \mbox{ else }.\\
\end{array}
\right.
\end{equation}

For example, the truncated matrix $ A^{(6)}$ has the form
\begin{equation}\label{F3}
 A^{(6)}=\left(
\begin{array}{cccccc}
 \omega ^2-\omega_0^2 & \frac{F}{2} & 0 & 0 & 0 & 0 \\
 -\frac{F \omega }{2} & -2 \omega  & -\frac{3 F \omega }{2} & 0 & 0 & 0 \\
 0 & \frac{F}{2} & 9 \omega ^2-\omega_0^2 & \frac{F}{2} & 0 & 0 \\
 0 & 0 & -\frac{3 F \omega }{2} & -4 \omega  & -\frac{5 F \omega }{2} & 0 \\
 0 & 0 & 0 & \frac{F}{2} & 25 \omega ^2-\omega_0^2 & \frac{F}{2} \\
 0 & 0 & 0 & 0 & -\frac{5 F \omega }{2} & -6 \omega  \\
\end{array}
\right)\;.
\end{equation}
 The truncated system of linear equations of the form
$A^{(N)}\,{\mathbf x}={\mathbf f}$ has the formal solution ${\mathbf x}=-F\,z_0\,\left(A^{(N)}\right)^{-1}_1$, where $\left(A^{(N)}\right)^{-1}_1$ denotes
the first column of the inverse matrix of $A^{(N)}$. Fortunately, there exists a recursion formula for the inverse of tridiagonal
matrices in terms of leading principal minors and co-leading principal minors, see \cite{U94}. Recall that a leading minor
of $A^{(N)}$ of order $n$ is the determinant of the submatrix of matrix elements in rows and columns from $1$ to $n$. Similarly, we will denote
by the ``co-leading principal minor $\phi_n$ of co-order $n$"
the determinant of the submatrix of matrix elements of $A^{(N)}$ in rows and columns from $n$ to $N$.
Since we do not need the whole matrix $\left(A^{(N)}\right)^{-1}$ but only its first column it turns out that only co-leading principal minors are involved.
It is well-known that the determinant of a tridiagonal matrix satisfies a two-term recursion relation. For our problem this implies the following system of recursion relations for
the $\phi_n,\;n=N,N-1,\ldots,1$.
\begin{eqnarray}
\label{F4a}
  \phi_N &=& A_{N,N}=\left\{
\begin{array}{l@{\;:\;}l}
 -N\,\omega & N \mbox{ even },\\
\left( N\,\omega\right)^2-\omega_0^2
 &  N \mbox{ odd },
\end{array}
\right.
\end{eqnarray}
\begin{equation}
\phi_{N-1}= A_{N-1,N-1}\,A_{N,N}-A_{N,N-1}\,A_{N-1,N}
\end{equation}
\begin{equation}
 = \left\{
\begin{array}{l@{\;:\;}l}
 \frac{1}{4} F^2 \omega  (N-1)-\omega  N \left(\omega ^2 (N-1)^2-\omega_0^2\right)
 & N \mbox{ even},\\
\frac{1}{4} F^2 \omega  N-\omega  (N-1) \left(\omega ^2 N^2-\omega_0^2\right)
&  N \mbox{ odd},
\end{array}
\right.
\end{equation}
\begin{equation}\label{F4d}
\phi_n = \left\{
\begin{array}{l@{\;:\;}l}
 -n\, \omega\,\phi_{n+1}+\frac{n+1}{4}F^2 \omega\,\phi_{n+2}
 & n \mbox{ even},\\
\left((n\,\omega)^2-\omega_0^2\right)\,\phi_{n+1}+\frac{n}{4}F^2 \omega\,\phi_{n+2}
&  n \mbox{ odd},
\end{array}
\right.
\end{equation}
Especially, $\phi_1=\det A^{(N)}$. Then the first column of $\left(A^{(N)}\right)^{-1}$ can be expressed
in terms of the co-leading principal minors and certain products of the lower secondary diagonal elements of $A$,
see the theorem in \cite{U94} for the special case of $j=1$. We write down the corresponding result for the Fourier
coefficients $x_n$:
\begin{eqnarray}
\label{F5a}
  x_1 &=& -F\,z_0\,\phi_2/\phi_1\;, \\
  \nonumber
  x_n &=&\left\{
\begin{array}{l@{\;:\;}l}
z_0 \frac{(-1)^n 2 \left(\frac{F}{2}\right)^n (-\omega )^{n/2} (n-1){!!}\,\phi_{n+1}}{\phi_1}
 & n \mbox{ even},\\
z_0 \frac{(-1)^n 2 \left(\frac{F}{2}\right)^n (-\omega )^{\frac{n-1}{2}} (n-2){!!}\,\phi _{n+1}}{\phi_1}
&  n \mbox{ odd},
\end{array}
\right.\\
&&\label{F5b}
\end{eqnarray}
where (\ref{F5b}) holds for $n=2,3,\ldots,N$.
$z_0$ is a free parameter that necessarily occurs due to the fact that the Fourier series solution is not yet normalized.
It could be chosen as $z_0=1$ or, alternatively, as $z_0=\phi_1$.
Depending on the context both choices will be adopted in what follows.
The latter choice has the following advantage: If $\phi_1=0$ the above solution
(\ref{F5a}) --  (\ref{F5b}) is no longer defined, but choosing $z_0=\phi_1$ and cancelling the fraction $z_0/\phi_1$ to $1$ we obtain a solution
that is always defined. Upon the choice $z_0=\phi_1$ the vanishing of $\phi_1=\det A^{(N)}$ is equivalent to the vanishing of the
time average $\overline{Z(t)}=z_0$. We recall the fact that this in turn characterizes the occurrence of resonances,
see Assertion \ref{ass2}.

\begin{figure}[ht]
  \centering
    \includegraphics[width=1.0\linewidth]{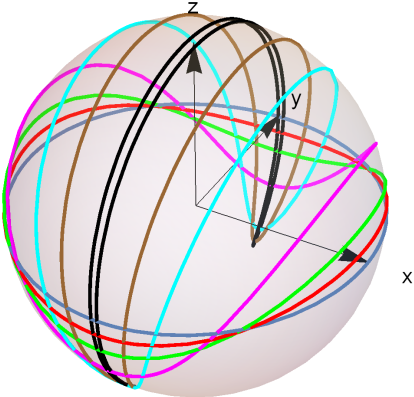}
  \caption[Q5]
  {Various periodic solutions of the classical RPL for $\omega=\omega_{res}^{(1)}(F)$ and $\omega_0=1$ visualized as closed trajectories on the Bloch sphere. The colored curves  correspond to different values of $F$, namely
  $F=1/10$ (blue), $F=1/2$ (red), $F=1$ (green), $F=2$ (magenta), $F=5$ (cyan), $F=10$ (brown), and $F=100$ (black).
  }
  \label{FIGQ5}
\end{figure}

At any case, from these recursion relations it is clear that each $x_n$ is a rational
function $\rho(n,N,F,\omega,\omega_0)$ in the variables $F,\omega$ and $\omega_0$. It can hence be viewed as a kind of Pad\'{e} approximation
for $x_n$ that becomes more and more exact for increasing $N$. For the choice $z_0=\phi_1$ the rational
function $\rho(n,N,F,\omega,\omega_0)$ becomes a polynomial in the variables  $F,\omega$ and $\omega_0$.

In order to give an impression of the structure of $\rho(n,N,F,\omega,\omega_0)$ we give the results for the $N=4$ truncation with $z_0=\phi_1$
although this will not yet be a good approximations of the exact RPL solutions:
\begin{eqnarray}
\nonumber
  \scriptstyle X(t) &=&  \scriptstyle \frac{1}{2} F \omega ^2 \omega _0 \left(9 F^2+16 \left(\omega _0^2
  -9 \omega^2\right)\right) \cos (\omega t )-F^3 \omega ^2 \omega _0\cos (3 \omega t ),\\
  && \label{F6a}\\
  \nonumber
  \scriptstyle Y(t)&=&  \scriptstyle \frac{1}{2} F \omega ^3 \left(9 F^2+16 \left(\omega _0^2-9 \omega ^2\right)\right) \sin ( \omega t )
 -3 F^3 \omega ^3 \sin (3  \omega t ),\\
   && \label{F6b}\\
   \nonumber
  \scriptstyle  Z(t)&=&  \scriptstyle \delta-\frac{1}{8} F^2 \omega ^2 \left(3 F^2+16 \left(\omega _0^2-9 \omega ^2\right)\right)\cos (2 \omega t )
  +\frac{3 F^4 \omega ^2}{8} \cos (4  \omega t),  \\
    && \label{F6c}\\
   \nonumber
  \scriptstyle \delta &=& \scriptstyle \frac{1}{16} \omega ^2 \left(3 F^4-8 F^2 \left(27 \omega ^2-11 \omega _0^2\right)+128
   \left(\omega ^2-\omega _0^2\right) \left(9 \omega ^2-\omega _0^2\right)\right)\\
   && \label{F6d}
\end{eqnarray}

In the Figure \ref{FIGQ5} we show solutions of the classical RPL for different $F$ at the resonance frequency  $\omega_{res}^{(1)}(F)$ that will be calculated in the next subsection. These solutions are based on the truncated Fourier series (\ref{F1a}) -- (\ref{F1c})  with $N=20$ and the choice $z_0=\phi_1=0$.
They can be either calculated by directly evaluating the Fourier series or by numerically solving the equation of motion with the initial values
obtained by the Fourier series. The differences between both methods are negligible. For small $F$ the solution approximately corresponds to a
uniform rotation in the $x-y$-plane, whereas for large $F$ the solution curve folds into a sort of pendulum motion in the $y-z$ plane that approaches the limit solution for $F,\omega\rightarrow\infty$, see subsection \ref{sec:LCOM0}.

%%%%%%%%%%%%%%%%%%%%%%%%%%%%%%%%%%%%%%%%%%%%%%%%%%%%%%%%%%%%%%%%%%%%%%%%%%%%%%%%%%%%%%%%%%%%%%%%%%%%%%%%%%%%%%%%%%%%%%%%%%%%%%%
\subsection{Calculation of the resonance frequencies}\label{sec:AARF}
%%%%%%%%%%%%%%%%%%%%%%%%%%%%%%%%%%%%%%%%%%%%%%%%%%%%%%%%%%%%%%%%%%%%%%%%%%%%%%%%%%%%%%%%%%%%%%%%%%%%%%%%%%%%%%%%%%%%%%%%%%%%%%%

\begin{figure}[ht]
  \centering
    \includegraphics[width=1.0\linewidth]{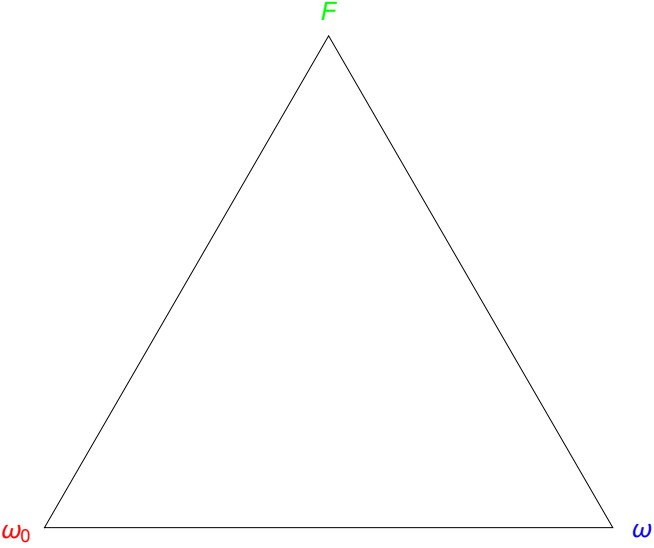}
  \caption[G3]
  {The domain of the three variables $\omega_0,\omega,F$ of the homogeneous function $\epsilon$ can be realized by an equilateral triangle $\triangle$. Each  point inside the triangle can be uniquely written as a convex sum of the vectors pointing to the three vertices, i.~e., as a sum with positive coefficients that add to unity. The three open edges represent limit cases, e.~g., the edge opposite to the vertex labelled ``$\omega$" represents the limit case $\omega\rightarrow 0$ such that $\omega_0$ and $F$ are kept finite. The three vertices themselves represent the limit cases where two of the three variables approach $0$ and the third one approaches $1$.
  }
  \label{FIGG3}
\end{figure}

\begin{figure}[ht]
  \centering
    \includegraphics[width=1.0\linewidth]{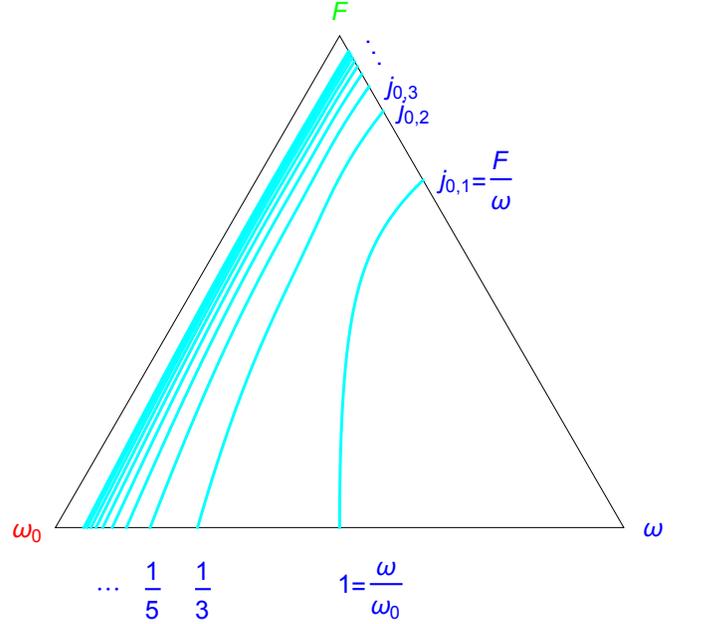}
  \caption[G4]
  {The first ten resonance curves ${\mathcal R}_n,\,n=1,\ldots, 10$ in the triangular domain of variables $\omega_0,\omega,F$. They intersect the line $F=0$ at $\frac{\omega}{\omega_0}=\frac{1}{2n-1}$ and the line $\omega_0=0$ at $\frac{F}{\omega}=j_{0,n}$ (the latter denoting the zeroes of the Bessel function $J_0$), as will be shown below, see (\ref{LCFR1}) and (\ref{V11}).
  }
  \label{FIGG4}
\end{figure}
We have shown in section \ref{sec:H} that $\epsilon$ is a positively homogeneous function and the same holds for its restriction to
the variables $\omega_0,\omega,F$ in the limit $G\rightarrow 0$.
The domain of these variables can be restricted to a two-dimensional domain without loss of information.
Instead of eliminating one of the three variables, which is inappropriate in some cases, one could introduce the scaled variables
\begin{equation}\label{H6}
  \widetilde{\omega_0}\equiv \frac{\omega_0}{\omega+\omega_0+F},
  \quad\widetilde{\omega}\equiv \frac{\omega}{\omega+\omega_0+F},
  \quad\widetilde{F}\equiv \frac{F}{\omega+\omega_0+F},
\end{equation}
that satisfy $0<\widetilde{\omega_0},\widetilde{\omega},\widetilde{F}<1$ and $\widetilde{\omega_0}+\widetilde{\omega}+\widetilde{F}=1$. The domain of the scaled variables
is an open equilateral triangle $\triangle$, see Figure \ref{FIGG3}. If the values of a positively homogeneous function like the quasienergy $\epsilon$
are known for arguments in $\triangle$ the function can be uniquely extended to the whole positive octant. If it is clear that we mean the scaled variables the tilde will be omitted.

The transformation between the three scaled variables $\widetilde{\omega_0},\widetilde{\omega},\widetilde{F}$
and the two Cartesian coordinates $x,y$ defining the points of $\triangle$ has the following form:
\begin{eqnarray}
\label{H7a}
  x &=&\frac{1}{2} \left( \widetilde{\omega}- \widetilde{\omega_0}\right)\;, \\
  \label{H7b}
  y &=& \frac{\sqrt{3}}{2}\widetilde{F}\;, \\
  \label{H7c}
     \widetilde{\omega_0} &=& \frac{1}{2}-x- \frac{y}{\sqrt{3}}\\
  \label{H7d}
 \widetilde{\omega} &=& \frac{1}{2}+x- \frac{y}{\sqrt{3}}\\
  \label{H7e}
 \widetilde{F} &=& \frac{2\,y}{\sqrt{3}}\;.
\end{eqnarray}

The resonance frequencies $\omega_{res}^{(n)},\;n=1,2,\ldots$ can be represented by ``resonance curves" ${\mathcal R}_n$ in the triangular domain
$\triangle$, see Figure \ref{FIGG4}. These curves have been numerically calculated by setting $z_0=\phi_1=\mbox{ det } A^{(50)}=0$,
see subsection \ref{sec:AATF} and Assertion \ref{ass2}. Later we will show that their intersections with the two edges $F=0$ and $\omega_0=0$ can be analytically determined, see Figure \ref{FIGG4}. For large $n$ the resonance curves approach the straight line segments determined by these intersections.

%%%%%%%%%%%%%%%%%%%%%%%%%%%%%%%%%%%%%%%%%%%%%%%%%%%%%%%%%%%%%%%%%%%%%%%%%%%%%%%%%%%%%%%%%%%%%%%%%%%%%%%%%%%%%%%%%%%%%%%%%%%%%%%
\subsection{Calculation of the quasienergy}\label{sec:AAQE}
%%%%%%%%%%%%%%%%%%%%%%%%%%%%%%%%%%%%%%%%%%%%%%%%%%%%%%%%%%%%%%%%%%%%%%%%%%%%%%%%%%%%%%%%%%%%%%%%%%%%%%%%%%%%%%%%%%%%%%%%%%%%%%%

After approximating $X(t),\,Y(t),\,Z(t)$ by a truncated Fourier series it is possible to calculate the quasienergy
$\epsilon=\frac{1}{2}\left(\omega_0+\overline{\left(\frac{F\cos(\omega t)\, X(t)}{R+Z(t)}\right)}\right)$ by a numerical integration. Alternatively,
we can determine the integral over $t\in[0,\frac{2\pi}{\omega}]$ analytically in the following way: We choose $F,\omega_0$ and $\omega$
as rational numbers and substitute $\cos(n\,\omega t)=\frac{1}{2}\left(z^n+\frac{1}{z^n}\right)$. This converts
$\frac{F\cos(\omega t)\, X(t)}{R+Z(t)}$ into a rational function of the complex variable $z$ and transforms the $t$-integral into
a contour integral over the unit circle in the complex plane that can be evaluated by means of the residue theorem.
The poles of the rational function to be integrated are obtained by the software packet MATHEMATICA as ``root objects" (this is the reason to choose
rational numbers for $F,\omega_0$ and $\omega$). The corresponding residues are exactly evaluated, mostly again in the form of root objects.

Both methods agree within the working precision but the numerical results are obtained faster. It may happen that the
quasienergy determined by these methods jumps into the ``wrong branch" and has to be corrected by adding an integer multiple of
$\omega$, see the corresponding discussion in subsection \ref{sec:GenSU2}.
In this way we obtain representations of the branches $\epsilon_i(F,\omega)$ without any restriction to the  values of $F$ and $\omega_0$.

\begin{figure}[ht]
  \centering
    \includegraphics[width=1.0\linewidth]{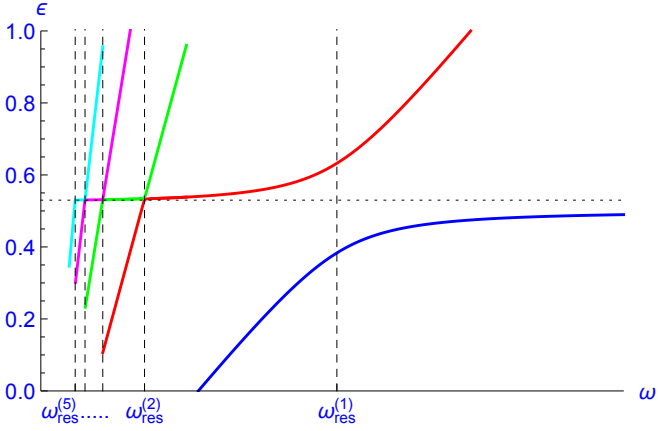}
  \caption[Q4]
  {The various branches of the quasienergy $\epsilon$ for $F=1/2$ and $\omega_0=1$ as functions of the frequency $\omega$
  calculated by an $N=20$ truncated Fourier series solution of the classical RPL.
  In the neighbourhood of the resonance frequencies $\omega_{res}^{(n)}$
  we have an avoided level crossing visualized by choosing the same color for the continuous branches of the quasienergy. For $n>1$ one has the impression of crossing levels due to the low resolution of the figure but, e.~g.~, the comparison with Figure \ref{FIGQ3} for $n=2$  shows that the levels actually do  \textit{not} cross. The dotted horizontal line indicates the value of $\lim_{\omega\rightarrow 0}\epsilon(\omega)\,\approx 0.52992$, see (\ref{F9}).
  }
  \label{FIGQ4}
\end{figure}

\begin{figure}[ht]
  \centering
    \includegraphics[width=1.0\linewidth]{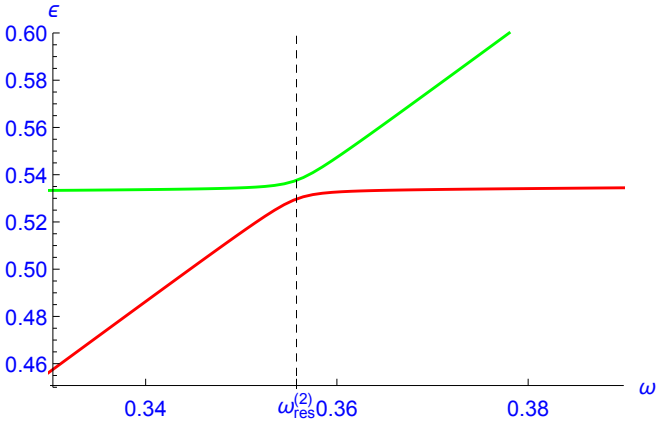}
  \caption[Q3]
  {The quasienergies $\epsilon_i,\;i=1,2$ for $F=1/2$ and $\omega_0=1$ as functions of the frequency $\omega$ in the neighbourhood of the
  resonance frequency $\omega_{res}^{(2)}\approx 0.355776$. $\epsilon_1$ (green curve) has been calculated analytically and numerically
  as described in the text and $\epsilon_2$ (red curve) has been chosen as  $\epsilon_2=3\,\omega-\epsilon_1$.
  }
  \label{FIGQ3}
\end{figure}

We have drawn a couple of branches of the function $\epsilon(\omega)$ for fixed $F$ and $\omega_0$, see Figurs \ref{FIGQ4}
and \ref{FIGQ3}. At the resonance frequencies $\omega_{res}^{(n)},\;n=1,\ldots,5$ that have been calculated according to the method
described in subsection \ref{sec:AARF} we observe avoided level crossings analogous to those obtained in the literature, see, e.~g.~,
\cite{AT55}, figure $1$ or \cite{S65}, figure $1$.\\

%%%%%%%%%%%%%%%%%%%%%%%%%%%%%%%%%%%%%%%%%%%%%%%%%%%%%%%%%%%%%%%%%%%%%%%%%%%%%%%%%%%%%%%%%%%%%%%%%%%%%%%%%%%%%%%%%%%%%%%%%%%%%%%
\section{Special limit cases}\label{sec:LC}
%%%%%%%%%%%%%%%%%%%%%%%%%%%%%%%%%%%%%%%%%%%%%%%%%%%%%%%%%%%%%%%%%%%%%%%%%%%%%%%%%%%%%%%%%%%%%%%%%%%%%%%%%%%%%%%%%%%%%%%%%%%%%%%

The introduction of $\triangle$ in subsection \ref{sec:AARF} as the natural domain of the arguments of
$\epsilon$ also clarifies the consideration of the various limit cases. We have three limit cases
where one of the scaled variables approaches $0$ but the other two variables remain finite. These three cases correspond to the three open edges of $\triangle$ and will be considered in the corresponding following subsections. First, the limit case $F\rightarrow 0$ is covered by a Fourier-Taylor series solution for ${\mathbf X}(t)$ and $\epsilon$, see subsection \ref{sec:LCF}.
The second case of  $\omega_0\rightarrow 0$ is considered in subsection \ref{sec:LCOM0} where we have calculated the asymptotic solution ${\mathbf X}(t)$ and the quasienergy $\epsilon$  up to linear terms in $\omega_0$.
It is very difficult to extend these results to higher orders of $\omega_0$ and hence we will contend ourselves with numerical approximations.
Finally, in the limit case $\omega\rightarrow 0$ we have recursively determined the terms of an $\omega$-power series for ${\mathbf X}(t)$ and explicitly calculated the first two terms of $\epsilon_{asy}=\epsilon_0+\epsilon_2\,\omega^2+O(\omega^4)$, see subsection \ref{sec:LCOM} .\\

There are three further ``limit cases of the limit cases" where two of the three scaled variables approach $0$ and the third one necessarily approaches $1$. They correspond to the three vertices of $\triangle$ and are not automatically included in the previous limit cases where we assumed
that one scaled variable approaches $0$ but the other two remain finite. Consider first the case where the unscaled variable $\omega$ approaches $\infty$ and the other two unscaled variables $F,\omega_0$ remain finite. Then, by (\ref{H6}), the scaled variables $\widetilde{F}$ and $\widetilde{\omega_0}$
will approach $0$ whereas $\widetilde{\omega}\rightarrow 1$. In this case we have calculated an FT series for ${\mathbf X}(t)$ in powers of
${\mathcal T}\equiv \frac{1}{\omega}$ and  the corresponding power series of the quasienergy,  see section \ref{sec:FQW}.

The next case of $\widetilde{F}, \widetilde{\omega}\rightarrow 0$, or, equivalently $\omega_0\rightarrow \infty$ can be treated either by considering the lowest order of $F$ in (\ref{IV3}) and (\ref{IV5a})  or  the lowest order of $\omega$ in (\ref{FTX1a}) and (\ref{FTX1b}).
It follows that both cases yield the same result, see  section \ref{sec:FQW}.

The last case $\widetilde{\omega},\widetilde{\omega_0}\rightarrow 0$ or, equivalently $F\rightarrow \infty$,
is somewhat subtle since the two limits cannot be interchanged, see the discussion in section \ref{sec:LCOM}.
It will not be treated in a separate subsection.

%%%%%%%%%%%%%%%%%%%%%%%%%%%%%%%%%%%%%%%%%%%%%%%%%%%%%%%%%%%%%%%%%%%%%%%%%%%%%%%%%%%%%%%%%%%%%%%%%%%%%%%%%%%%%%%%%%%%%%%%%%%%%%%
\subsection{Limit case $F\rightarrow 0$}\label{sec:LCF}
%%%%%%%%%%%%%%%%%%%%%%%%%%%%%%%%%%%%%%%%%%%%%%%%%%%%%%%%%%%%%%%%%%%%%%%%%%%%%%%%%%%%%%%%%%%%%%%%%%%%%%%%%%%%%%%%%%%%%%%%%%%%%%%

%%%%%%%%%%%%%%%%%%%%%%%%%%%%%%%%%%%%%%%%%%%%%%%%%%%%%%%%%%%%%%%%%%%%%%%%%%%%%%%%%%%%%%%%%%%%%%%%%%%%%%%%%%%%%%%%%%%%%%%%%%%%%%%
\subsubsection{Resonance frequencies}\label{sec:LCFR}
%%%%%%%%%%%%%%%%%%%%%%%%%%%%%%%%%%%%%%%%%%%%%%%%%%%%%%%%%%%%%%%%%%%%%%%%%%%%%%%%%%%%%%%%%%%%%%%%%%%%%%%%%%%%%%%%%%%%%%%%%%%%%%%

\begin{table}
\caption{{\label{tab1}}Table of the coefficients of the power series (\ref{LCFR2}) for the resonance frequencies $\omega^{(1)}_{res}$.}

\begin{center}\begin{tabular}{|c|c|}\hline
$2\,m$ &  $\sigma_{2m}^{(1)}$\\ \hline\hline
$2$&$\frac{1}{16}$\\
\hline
$4$ & $\frac{1}{1024}$\\
\hline
$6$ & $-\frac{35}{131072}$\\
\hline
$8$ & $\frac{103}{8388608}$\\
\hline
$10$ & $\frac{1873}{805306368}$\\
\hline
$12$ &$ -\frac{1577257}{3710851743744}$\\
\hline
$14$ &$ \frac{67429531}{17099604835172352}$\\
\hline
$16$ &$ \frac{304008125947}{39397489540237099008}$\\
\hline
\end{tabular}
\end{center}
\end{table}

\begin{table}
\caption{{\label{tab2}}Table of the coefficients of the power series (\ref{LCFR2}) for the resonance frequencies $\omega^{(2)}_{res}$.}

\begin{center}\begin{tabular}{|c|c|}\hline
$2\,m$ &  $\sigma_{2m}^{(2)}$\\ \hline\hline
$2$&$\frac{3}{32}$\\
\hline
$4$ & $-\frac{135}{8192}$\\
\hline
$6$ & $ \frac{2133}{1048576}$\\
\hline
$8$ & $ \frac{588789}{536870912}$\\
\hline
$10$ & $ -\frac{98579025}{68719476736}$\\
\hline
$12$ &$  \frac{19157942853}{17592186044416}$\\
\hline
\end{tabular}
\end{center}
\end{table}

\begin{table}
\caption{{\label{tab3}}Table of the coefficients of the power series (\ref{LCFR2}) for the resonance frequencies $\omega^{(3)}_{res}$.}

\begin{center}\begin{tabular}{|c|c|}\hline
$2\,m$ &  $\sigma_{2m}^{(3)}$\\ \hline\hline
$2$&$\frac{5}{96}$\\
\hline
$4$ & $-\frac{2125}{221184}$\\
\hline
$6$ & $ \frac{1146875}{254803968}$\\
\hline
$8$ & $ -\frac{3244765625}{1174136684544}$\\
\hline
$10$ & $\frac{2045715078125}{1352605460594688}$\\
\hline
$12$ &$-\frac{558332576171875}{1038800993736720384}$\\
\hline
\end{tabular}
\end{center}
\end{table}

\begin{table}
\caption{{\label{tab4}}Table of the coefficients $\sigma_{2m}^{(n)}$
of the power series (\ref{LCFR2}) for the resonance frequencies $\omega^{(n)}_{res}$ for $m=1,2,3$ and $n=1,\ldots 10$.}
\begin{center}\begin{tabular}{|c|c|c|c|}\hline
$n$ &  $\sigma_{2}^{(n)}$&  $\sigma_{4}^{(n)}$ &$\sigma_{6}^{(n)} $\\ \hline\hline
$1$&$\frac{1}{16}$ & $\frac{1}{1024}$ & $ -\frac{35}{131072} $\\
\hline
$2$ & $\frac{3}{32}$&$-\frac{135}{8192}$ & $  \frac{2133}{1048576}$\\
\hline
$3$ & $ \frac{5}{96}$&$-\frac{2125}{221184}$ & $\frac{1146875}{254803968} $\\
\hline
$4$ & $ \frac{7}{192}$&$-\frac{12005}{1769472}$ & $\frac{120892751}{40768634880} $\\
\hline
$5$ & $\frac{9}{320}$&$-\frac{43011}{8192000}$ & $\frac{235598949}{104857600000} $\\
\hline
$6$ &$\frac{11}{480}$&$-\frac{118459}{27648000}$ & $\frac{10123182707}{5573836800000} $\\
\hline
$7$ &$\frac{13}{672}$&$-\frac{274625}{75866112}$ & $ \frac{32687521841}{21412451450880} $\\
\hline
$8$ &$\frac{15}{896}$&$-\frac{563625}{179830784}$ & $\frac{23778534375}{18046378835968}  $\\
\hline
$9$ &$\frac{17}{1152}$&$-\frac{1056295}{382205952}$ & $ \frac{2573069114971}{2219118333788160}$\\
\hline
$10$ &$\frac{19}{1440}$&$-\frac{1845071}{746496000}$ & $  \frac{2204002956989}{2128409395200000}$\\
\hline
\end{tabular}
\end{center}
\end{table}

A glimpse of (\ref{F3}) shows that for $F=0$ the determinant of $A^{(N)}$ vanishes for $\omega=\frac{\omega_0}{2n+1},\;n=1,2,\ldots$.
Hence the resonance condition $z_0=\phi_1=\det A^{(N)}=0$ for $N$ arbitrarily large, see Assertion \ref{ass2}, leads to
\begin{equation}\label{LCFR1}
 \omega_{res}^{(n)}=\frac{\omega_0}{2n-1},\;\mbox{ for }n=1,2,\ldots\mbox{ and }F=0
 \;.
\end{equation}
This explains the intersections of the resonance curves ${\mathcal R}_n$ with the edge $F=0$ of $\triangle$, see Figure \ref{FIGG4}.\\

By an analogous reasoning we may also calculate the first terms of the power series w.~r.~t.~$F$ of $\omega_{res}^{(n)}$ for small $n$
or small $m$. The power series has the form:
\begin{equation}\label{LCFR2}
 \omega_{res}^{(n)}=\frac{\omega_0}{2n-1}+ \sum_{m=1}^{\infty} \sigma_{2m}^{(n)}\,\omega_0^{1-2m}\, F^{2m}\;.
\end{equation}
Recall that the differences $\omega_{res}^{(n)}-\frac{\omega_0}{2n-1}$ are traditionally called ``Bloch-Siegert shifts".
The coefficients $\sigma_{2m}^{(n)}$ of (\ref{LCFR2}) can be determined as follows: We insert the power series (\ref{LCFR2})
into the expression of $\det A^{(N)}$ (for a suitable large $N$) and set the first few  coefficients of the resulting power
series w.~r.~t.~$F$ to zero. This yields recursive equations from which the $\sigma_{2m}^{(n)}$ may be determined, independent of $N$.
The corresponding results for $n=1,2,3$  are contained in the Tables \ref{tab1}, \ref{tab2}, \ref{tab3}. They are in accordance with
the three coefficients for $n=1$ published in \cite{S65} and with the results of \cite{HPS73} and \cite{AB74}.
Table \ref{tab4} contains the first  non-vanishing coefficients
$\sigma_{2m}^{(n)}$ for $n=1,\ldots,10$ and $m=1,2,3$ that have been determined in the same way.
A closed formula for some of the $\sigma_{2m}^{(n)}$ can be obtained by ``computer algebraic induction":
\begin{equation}\label{LCFR3}
  \sigma _2^{(n)}=\frac{2n-1}{2^4 (n-1) n} \mbox{  for } n>1
  \;,
\end{equation}
\begin{equation}\label{LCFR4}
  \sigma _4^{(n)}=-\frac{(2 n-1)^3 \left(3 (2 n-1)^2-7\right)}{2^{12} (n-1)^3 n^3} \mbox{  for } n>1
  \;,
\end{equation}
\begin{eqnarray}
\nonumber
    \sigma _6^{(n)} &=&\frac{(2 n-1)^5 }{2^{20}   (n-2) (n-1)^5 n^5 (n+1)}\,\times\\
   \nonumber
   && \left(5 (2 n-1)^6-57 (2 n-1)^4+187 (2 n-1)^2-199\right)\\
    \label{LCFR5}
   && \mbox{for   }n>2\;.
\end{eqnarray}

We find that the general coefficients $\sigma _{2k}^{(n)}$ can be written in the form
\begin{eqnarray}
\label{LCFR6a}
  \sigma _{2k}^{(n)} &=& (2n-1)^{2k-1} \frac{Z(k,n)}{N(k,n)},\quad\mbox{ where }\\
 Z(k,n) &=&\sum_{\mu=0}^{z(k)}A_\mu^{(k)}\,\left(n(n-1)\right)^\mu \;,\\
 N(k,n)&=&\prod _{j=1}^{2 \left\lceil \frac{k}{2}\right\rceil }
 \left(-2 \left\lceil\frac{k}{2}\right\rceil + 2 (j-1)+2n\right)^{n(k,j)}, \\
 n(k,j)&=&{2 \left\lfloor \frac{k}{2\left| 2 j-2 \left\lceil \frac{k}{2}\right\rceil -1\right|}\right\rfloor -1},\\
 z(k)&=& \frac{1}{2}\left(\sum _{j=1}^{2 \left\lceil \frac{k}{2}\right\rceil } n(k,j)\right)-k\;.
\end{eqnarray}

\begin{figure}[ht]
  \centering
    \includegraphics[width=1.0\linewidth]{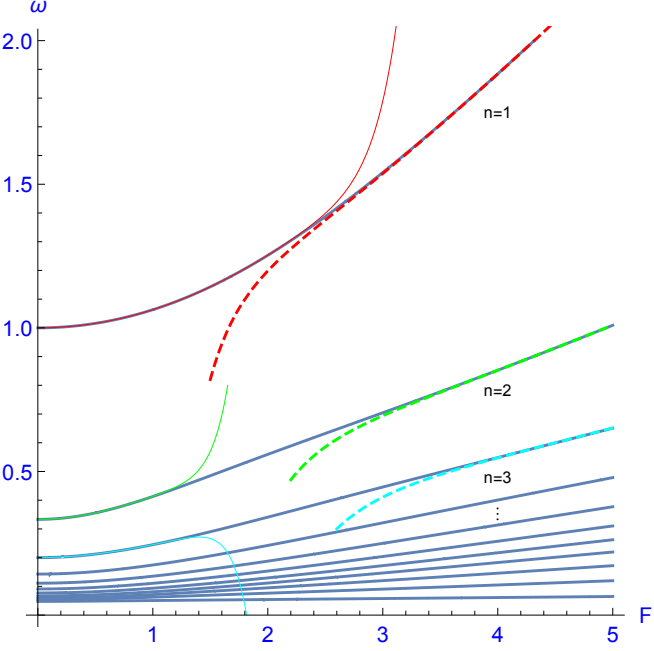}
  \caption[BS1]
  {Various resonance frequencies $\omega_{res}^{(n)}$ (blue curves) for $\omega_0=1$ as functions of $F$ for $n=1,2,\ldots,$ numerically determined
  by the resonance condition $\det A^{(20)}=0$.
  The thin continuous red/green/cyan curves correspond to the power series (\ref{LCFR2}) according to the results of Tables \ref{tab1} - \ref{tab3} for $n=1,2,3$.
   The corresponding dashed curves are the asymptotic four terms limits of  $\omega_{res}^{(n)}$ for $\omega,\,F\rightarrow\infty$
  according to (\ref{V12a}) -- (\ref{V12c}).
  }
  \label{FIGQ6}
\end{figure}

Figure \ref{FIGQ6} shows a couple of resonance curves $\omega_{res}^{(n)}$ as functions of $F$ together with their $F$-expansions.
The  numerical agreement between both curves is excellent but limited to bounded values of $F$. This clearly indicates a finite
radius of convergence for the power series (\ref{LCFR2}).
\\

%%%%%%%%%%%%%%%%%%%%%%%%%%%%%%%%%%%%%%%%%%%%%%%%%%%%%%%%%%%%%%%%%%%%%%%%%%%%%%%%%%%%%%%%%%%%%%%%%%%%%%%%%%%%%%%%%%%%%%%%%%%%%%%
\subsubsection{Fourier-Taylor series}\label{sec:FT}
%%%%%%%%%%%%%%%%%%%%%%%%%%%%%%%%%%%%%%%%%%%%%%%%%%%%%%%%%%%%%%%%%%%%%%%%%%%%%%%%%%%%%%%%%%%%%%%%%%%%%%%%%%%%%%%%%%%%%%%%%%%%%%

In the section \ref{sec:FTX}  we will present a solution of the classical RPL in terms of so-called Fourier-Taylor (FT) series.
A few explanations will be in order.
An FT series is a Taylor series of a
(vector) quantity ${\mathbf A}(F,t)$, periodic in $t$, w.~r.~t.~the parameter $F$ such that
each coefficient of $F^n$ is a finite Fourier series w.~r.~t.~the time
variable $t$ of maximal order $n$:
\begin{equation}\label{FT1}
{\mathbf A}(F,t)=\sum_{n=0}^{\infty}F^n\,\sum_{m=-n}^{n}\,{\mathbf A}_{nm}\,e^{{\sf i}\,m\,\omega\,t}\;.
\end{equation}
Put differently, the $T=\frac{2\pi}{\omega}$-periodic function ${\mathbf A}(F,t)$ is expanded into a Fourier series such
that each Fourier coefficient of order $m$ is a Taylor series w.~r.~t.~$F$ that starts with the lowest order of $n=m$.
Fourier series with components that are in turn Laurent series of a suitable parameter
are know as ``Poisson series" in celestial mechanics, see, e.~g., \cite{SA01}.
FT series are special Poisson series characterized by the restriction $\sum_{m=-n}^{n}$ in (\ref{FT1}) and have been applied in \cite{SB15}
to a couple of physical problems by utilizing computer algebraic means. It is possible to consider the more general case where the frequency
$\omega$ is not given but also calculated iteratively in terms of a Taylor series, but this generalization is not needed in the present context of RPL.

It is obvious that sums and products of FT series are again FT series. More generally, the power series of an FT series is again an FT series,
at least in the sense of formal power series. For practical applications the size of the convergence radius becomes important.

%%%%%%%%%%%%%%%%%%%%%%%%%%%%%%%%%%%%%%%%%%%%%%%%%%%%%%%%%%%%%%%%%%%%%%%%%%%%%%%%%%%%%%%%%%%%%%%%%%%%%%%%%%%%%%%%%%%%%%%%%%%%%%%
\subsubsection{FT series for ${\mathbf X}(t)$ and $\epsilon$}\label{sec:FTX}
%%%%%%%%%%%%%%%%%%%%%%%%%%%%%%%%%%%%%%%%%%%%%%%%%%%%%%%%%%%%%%%%%%%%%%%%%%%%%%%%%%%%%%%%%%%%%%%%%%%%%%%%%%%%%%%%%%%%%%%%%%%%%%

In the case $F=0$ there are only two normalized solutions of the classical RPL
that are  $T$-periodic for all $T>0$,
namely ${\mathbf X}(t)=\pm (0,0,1)^\top$. Hence for infinitesimal $F$ we expect that we still have
$Z(t)=\pm 1+O(F^2)$ but $(X(t),Y(t))$ will describe an infinitesimal ellipse, i.~e.~, $X(t)=F A \cos\omega t+O(F^3)$ and $Y(t)=F B \sin\omega t+O(F^3)$.
These considerations and numerical investigations suggest the following FT series ansatz:
\begin{eqnarray}
\nonumber
  X(t)&=&\sum_{n=0}^{\infty}F^{2n+1}\sum_{m=0}^{n}R_{n,m}(\omega,\omega_0)\cos(2m+1)\omega t\;, \\
  &&\label{FL2a}\\
 \nonumber
  Y(t) &=& \frac{1}{\omega_0}\sum_{n=0}^{\infty}F^{2n+1}\quad \times\\
  \nonumber
  &&\sum_{m=0}^{n}(2m+1)\omega\,R_{n,m}(\omega,\omega_0)\sin(2m+1)\omega t\;,\\
   \label{FL2b}
  &&  \\
  \label{FL2c}
  Z(t)&=& \sum_{n=0}^{\infty}F^{2n}\sum_{m=0}^{n}S_{n,m}(\omega,\omega_0)\cos 2m\omega t\;.
\end{eqnarray}
In the ansatz (\ref{FL2b}) for $Y(t)$ we have already used that $Y(t)$ is completely determined via
$Y(t)=-\frac{1}{\omega_0}\frac{d}{dt}X(t)$  and hence need not be further considered.
The other two differential  equations (\ref{AA1b}) and (\ref{AA1c}) yield recursion relations
for the functions $R_{n,m}(\omega,\omega_0)$ and $S_{n,m}(\omega,\omega_0)$. As  initial
conditions we impose the following choices:
\begin{eqnarray}
\label{FL3a}
  S_{0,0}(\omega,\omega_0) &=& 1\;,\\
    \label{FL3b}
  R_{0,0}(\omega,\omega_0)&=& -\frac{\omega_0}{(\omega -\omega_0) (\omega +\omega_0)}\;,\\
   \label{FL3c}
  S_{n,0}(\omega,\omega_0)&=& 0\;, \mbox{ for } n=1,2,\ldots\;.
\end{eqnarray}

For $n>0$ the FT coefficients $R_{n,m}$ and $S_{n,m}$ can be recursively determined by means of the
following relations:
\begin{eqnarray}
\nonumber
  S_{n,m} &=& -\frac{1}{4m\,\omega_0}\left((2m+1) R_{n-1,m} +(2m-1)R_{n-1,m-1}\right)\\
  \label{FL4a}
  && \mbox{ for } 1\le m\le n\;, \\
  \nonumber
  R_{n,m} &=& -\frac{\omega_0}{2\left( (2m+1)^2\omega^2-\omega_0^2\right)}\left(S_{n,m}+S_{n,m+1}\right)\\
   \label{FL4b}
   && \mbox{ for } 0\le m\le n\;,
\end{eqnarray}
where, of course, we have to set $S_{n,n+1}=0$ in (\ref{FL4b}). It follows
that $R_{n,m}(\omega,\omega_0)$ and $S_{n,m}(\omega,\omega_0)$ are rational functions of $\omega$ and $\omega_0$.

We recall that under the transformation (\ref{H2a}) -- (\ref{H2e}),  $X,Y$ and $Z$ remain invariant which entails
$R_{nm}\mapsto \lambda^{-2n-1 }\,R_{nm}$ and $S_{nm}\mapsto \lambda^{-2n}\, S_{nm}$. Then it easily follows that both sides of the
recursion relations (\ref{FL4a}) and (\ref{FL4b}) transform in the same way, which can be viewed as a consistency check of the ansatz
(\ref{FL2a}) -- (\ref{FL2c}).

We will show the first few terms of the FT series for $X(t)$ and $Z(t)$:
\begin{eqnarray}
\nonumber
  X(t) &=& -F\frac{\omega_0}{\omega^2-\omega_0^2}\,\cos\omega t\quad + \\
  \nonumber
   &&F^3\left( -\frac{\omega_0}{8 \left(\omega^2-\omega_0^2\right)^2}\,\cos\omega t\right.\\
  \nonumber
  &&\left. -\frac{\omega_0}{8 \left(9 \omega ^2-\omega_0^2\right)    \left(\omega^2-\omega_0^2\right)}\,
   \cos 3\omega t
   \right) + O(F^5),
    \\
    \label{FTX1a}
   &&\\
   \nonumber
   Z(t)&=& 1+F^2\,\frac{1}{4(\omega^2-\omega_0^2)}\cos 2\omega t\quad +\\
   \nonumber
   && F^4\left(
   \frac{3 \omega ^2-\omega _0^2}{8 \left(\omega ^2-\omega _0^2\right){}^2 \left(9 \omega ^2-\omega _0^2\right)}\cos 2\omega t\right.\\
   \nonumber
 && \left. +\frac{3}{64 \left(\omega ^2-\omega_0^2\right) \left(9 \omega ^2-\omega_0^2\right)}\cos 4\omega t \right)+O(F^6)\;.\\
 &&  \label{FTX1b}
\end{eqnarray}
We note that the coefficients
contain denominators of the form $\omega^2-\omega_0^2$ and $9\omega^2-\omega_0^2$ due to the denominator
$(2m+1)^2\omega^2-\omega_0^2$ in the recursion relation (\ref{FL4b}). Hence the FT series breaks down at the
resonance frequencies $\omega_{res}^{(m)}=\frac{\omega_0}{2m-1}$. This is the more plausible since
according to the above ansatz $z_0=1$ which is not compatible with the resonance condition $z_0=0$, see Assertion \ref{ass2}.

Using the FT series solution (\ref{FL2a}) -- (\ref{FL2c}) it is a straightforward task to calculate the quasienergy
$\epsilon=a_0$ as the time-independent part of the FT series of (\ref{C13d}) which in the present case of RPL assumes the form
\begin{equation}\label{FQ1}
\frac{1}{2}\left(\omega_0+\frac{F \cos(\omega t)X(t)}{R+Z(t)}\right)=a_0+\sum_{\stackrel{n\in{\mathbbm Z}}{n\neq 0}}a_n\,e^{{\sf i}\,n\,\omega\, t}\;.
\end{equation}
The first few terms of the result are given by
\begin{eqnarray}
\nonumber
 \epsilon &=& \frac{\omega_0}{2}-\frac{F^2\,\omega_0}{8 \left(\omega ^2-\omega_0^2\right)}+\frac{F^4\,\omega_0 \left(\omega ^2+3\omega_0^2\right)}
 {128 \left(\omega^2-\omega_0^2\right)^3} \\
  \nonumber
   &-&\frac{F^6\,\omega_0 \left(-5\omega_0^6+35 \omega ^2\omega_0^4+33 \omega ^4\omega_0^2+\omega ^6\right)}
   {512 \left(\omega ^2-\omega_0^2\right)^5\left(9 \omega ^2-\omega_0^2\right)}+ O(F^8)\;.\\
    \label{FQ2}
   &&
\end{eqnarray}
This is in agreement with \cite{S65}, equation (29), except for the first term which is probably a typo.
\\

It will be instructive to check the first two terms of (\ref{FQ2}) by using the decomposition of the quasienergy into
a dynamical and a geometrical part in section \ref{sec:GeoI}. In lowest order in $F$ the classical RPL solution is a motion on
an ellipse with semi axes $a=\frac{F\,\omega_0}{|\omega^2-\omega_0^2|}$ and $b=\frac{F\,\omega}{|\omega^2-\omega_0^2|}$.
Hence the geometrical part of the quasienergy reads
\begin{equation}\label{FQ4}
 \epsilon_g=\frac{\omega}{4\pi}\,\pi\,a\,b+O\left(F^4\right) = \frac{F^2 \omega ^2\omega_0}{4 \left(\omega ^2-\omega_0^2\right)^2}+O\left(F^4\right)\;.
\end{equation}
The dynamical part is obtained as
\begin{equation}\label{FQ5}
   \epsilon_d=\overline{\frac{\omega_0\,Z+X\, F \,\cos\omega t}{2 R}}
   =\frac{\omega_0}{2}
   +\frac{\omega _0 \left(\omega _0^2-3 \omega ^2\right)}{8 \left(\omega ^2-\omega_0^2\right)^2}F^2+O\left(F^4\right).
\end{equation}
The sum of both parts together correctly yields
\begin{equation}\label{FQ6}
\epsilon=\epsilon_d + \epsilon_g= \frac{\omega_0}{2}-\frac{F^2\,\omega_0}{8( \omega ^2-\omega_0^2)}+O\left(F^4\right)\;.
\end{equation}
Moreover,
\begin{equation}\label{FQ7}
\frac{\partial \epsilon}{\partial \omega} =\frac{ \omega\,  \omega _0}{4 \left(\omega ^2-\omega _0^2\right)^2}\,F^2 +O(F^4)=\frac{\epsilon_g}{\omega}
\;,
\end{equation}
in accordance with Assertion \ref{ass3}.

However, it is plausible from (\ref{FQ2}) that the FT series for the quasienergy has poles at the values $\omega=\omega_{res}^{(m)} = \frac{1}{2m-1},\;m=1,2,\ldots$
and hence the present FT series ansatz is not suited to investigate the Bloch-Siegert shift for small $F$.
We have also found a modified FT series that is valid in the neighbourhood of $\omega_{res}^{(1)}$ but will not dwell upon this.

%%%%%%%%%%%%%%%%%%%%%%%%%%%%%%%%%%%%%%%%%%%%%%%%%%%%%%%%%%%%%%%%%%%%%%%%%%%%%%%%%%%%%%%%%%%%%%%%%%%%%%%%%%%%%%%%%%%%%%%%%%%%%%%
\subsection{Limit case $\omega_0\rightarrow 0$}\label{sec:LCOM0}
%%%%%%%%%%%%%%%%%%%%%%%%%%%%%%%%%%%%%%%%%%%%%%%%%%%%%%%%%%%%%%%%%%%%%%%%%%%%%%%%%%%%%%%%%%%%%%%%%%%%%%%%%%%%%%%%%%%%%%%%%%%%%%%

%%%%%%%%%%%%%%%%%%%%%%%%%%%%%%%%%%%%%%%%%%%%%%%%%%%%%%%%%%%%%%%%%%%%%%%%%%%%%%%%%%%%%%%%%%%%%%%%%%%%%%%%%%%%%%%%%%%%%%%%%%%%%%%
\subsubsection{The classical equation of motion}\label{sec:CE}
%%%%%%%%%%%%%%%%%%%%%%%%%%%%%%%%%%%%%%%%%%%%%%%%%%%%%%%%%%%%%%%%%%%%%%%%%%%%%%%%%%%%%%%%%%%%%%%%%%%%%%%%%%%%%%%%%%%%%%%%%%%%%%%

We reconsider the classical RPL equation of motion (\ref{AA1a}) -- (\ref{AA1c})
and look for solutions that are at most linear in $\omega_0$, neglecting higher order terms.
For $\omega_0=0$ we have the exact ``pendulum solution"
\begin{eqnarray}
\label{V2a}
  X(t) &=& 0 \;,\\
 \nonumber
  Y(t) &=& -\sin\left( f\, \sin\omega t\right)= -2\,\sum_{n=1,3,\ldots} J_n(f)\,\sin n\omega t\;,\\
  && \label{V2b}\\
\nonumber
   Z(t)&=&\cos\left( f\, \sin\omega t\right)= J_0(f)+2\,\sum_{n=2,4,\ldots} J_n(f)\,\cos n\omega t\;.\\
  &&\label{V2c}
\end{eqnarray}
Here the $J_n(\ldots)$ denote the Bessel functions of first kind and integer order
and we have set $f\equiv \frac{F}{\omega}$ since this combination permanently occurs in what follows.
(\ref{V2b}) and (\ref{V2c}) follow from the Jacobi-Anger expansion.
Moreover, we will assume
\begin{equation}\label{V2d}
  0<f<\pi
  \;,
\end{equation}
in order to avoid problems with the following integrations.
We note that if a constant $x$-component $X(t)=x_0$  would be added to the above solution it would still solve
(\ref{AA1a}) -- (\ref{AA1c}) for $\omega_0=0$. But only the choice $X(t)=x_0=0$ is suited as a starting point for higher orders of $\omega_0$.

The next, linear order of the solution of  (\ref{AA1a}) -- (\ref{AA1c}) is obtained by replacing (\ref{V2a}) by
\begin{eqnarray}
\label{V3a}
  X(t) &=& \omega_0\,X_1(t),\mbox{ where} \\
  \label{V3b}
  \dot{X}_1(t) &=&-Y(t)= \sin\left( f\, \sin\omega t\right)\;.
\end{eqnarray}
A periodic solution of (\ref{V3b}) is given by the Fourier series
\begin{equation}\label{V4}
 X_1(t)=-\frac{2}{\omega}\sum_{n=1,3,\ldots}\frac{1}{n}\,J_n(f)\,\cos n\omega t
 \;.
\end{equation}
The radius of the Bloch sphere for this solution is still $R=1+O\left(\omega_0^2\right)$.

We want to determine the quasienergy $\epsilon$ in linear order in $\omega_0$ which according to (\ref{C13d}) reads
\begin{eqnarray}\label{V5a}
\epsilon&=&\frac{1}{2}\left( \omega_0+\overline{\left(\frac{h_1\, X}{1+Z} \right)} \right)\\
\label{V5b}
&\stackrel{(\ref{V3a})}{=}&\frac{\omega_0}{2}\left( 1+\overline{\left(\frac{h_1\, X_1}{1+Z} \right)} \right)\;,
\end{eqnarray}
where $h_1\equiv F\,\cos\omega t$.
For the time average we make the substitution $\tau = \omega\,t$ and perform the $\tau$-integration over the interval $[0,2\pi]$.
This yields a factor $1/\omega$ for each time integral which is partially compensated by the factor $\frac{\omega}{2\pi}$ due to
the time average.
Then the time average integral in (\ref{V5b}) can be transformed by partial integration into
\begin{equation}\label{V6}
\int_{0}^{2\pi}X_1\;\frac{h_1}{1+Z}\;d\tau \equiv \int_{0}^{2\pi}u\; \frac{dv}{d\tau}\;d\tau=\left[u\,v\right]_0^{2\pi}- \int_{0}^{2\pi}\frac{du}{d\tau}\; {v}\;d\tau
\;.
\end{equation}
By (\ref{V3b}) we have
\begin{equation}\label{V6a}
\frac{du}{d\tau}=\frac{1}{\omega}\frac{du}{d t} = \frac{1}{\omega} \dot{X}_1= \frac{1}{\omega}\sin\left( f\, \sin \tau\right)\;.
\end{equation}
In order to calculate $v$ we consider the integral
\begin{eqnarray}
\label{V7a}
  v &=& \int \frac{h_1}{1+Z}\,d\tau \\
  \label{V7b}
   &=& F\,\int \frac{\cos \tau}{1+\cos(f \sin \tau)}\,d\tau\;.
\end{eqnarray}
Substituting $x=f\sin \tau$, hence $dx= f\,\cos \tau\;d\tau$, we obtain
\begin{eqnarray}
\label{V8a}
  v &=&\omega\, \int \frac{1}{1+\cos x}\,dx \\
  \label{V8b}
   &=&\omega\,\tan\frac{x}{2}= \omega\,\tan\left(\frac{f}{2}\sin \tau\right)\;,
  \end{eqnarray}
suppressing irrelevant integration constants. Since $u$ and $v$ are $2\pi$-periodic functions the term $\left[u\,v\right]_0^{2\pi}$ in (\ref{V6}) vanishes. By (\ref{V6a}) and (\ref{V8b})
the remaining integral reads
\begin{eqnarray}
\nonumber
 - \int_{0}^{2\pi}\frac{du}{d\tau}\; {v}\;d\tau &=& -\int_{0}^{2\pi}\sin\left( f \sin \tau\right)\,\tan\left(\frac{f}{2}\sin \tau\right)\;d\tau \\
 && \label{V9a}\\
 \label{V9b}
 &=& - \int_{0}^{2\pi}\left(1-\cos\left(f \sin \tau\right)\right)\,d\tau\\
 \label{V9c}
 &=& = -2\pi\,\left( 1-J_0(f)\right)\;,
\end{eqnarray}
using (\ref{V2c}) in the last step.
After dividing by $2\pi$ due to the $\tau$-average we obtain for (\ref{V5b}):
\begin{equation}\label{V10}
 \epsilon= \frac{\omega_0}{2}J_0\left(\frac{F}{\omega}\right)
 \;.
\end{equation}
The decomposition into dynamical and geometrical part of the quasienergy according to section \ref{sec:GeoI} reads
\begin{eqnarray}
\label{V10a}
  \epsilon_g &=& \frac{\omega_0}{2}\left(\frac{F}{\omega}J_1\left(\frac{F}{\omega}\right)\right)\;, \\
  \label{V10b}
  \epsilon_d &=& \frac{\omega_0}{2}\left(J_0\left(\frac{F}{\omega}\right)-\frac{F}{\omega}J_1\left(\frac{F}{\omega}\right)\right)\;.
\end{eqnarray}
Note further that
\begin{equation}\label{V10c}
  \frac{\partial \epsilon}{\partial \omega}=\frac{F \omega _0}{2 \omega ^2} J_1\left(\frac{F}{\omega }\right)=\frac{\epsilon_g}{\omega}
\end{equation}
in accordance with Assertion \ref{ass3}.

Moreover, it is clear from (\ref{V10}) that the resonance condition $\frac{\partial \epsilon}{\partial \omega_0}=0$, cp.~(\ref{R2}),
is equivalent to
\begin{equation}\label{V11}
 \omega=\omega_{res}^{(n)}=\frac{F}{j_{n,0}}\;,
\end{equation}
where $j_{n,0}$ denotes the $n$-th zero of the Bessel function $J_0$. This yields the intersections of the resonance curves
${\mathcal R}_n$ with the line $\omega_0=0$, see Figure \ref{FIGG4}.
Note further that, by (\ref{V2c}),  $J_0\left(\frac{F}{\omega}\right)$ is the Fourier coefficient
of $Z(t)$ corresponding to the constant term and hence $\overline{Z(t)}=J_0\left(\frac{F}{\omega}\right)$ vanishes exactly in the
resonance case, in accordance with Assertion \ref{ass2}.

Unfortunately, the integrals occurring in the next, quadratic and cubic orders in $\omega_0$ cannot be solved in closed form and we cannot extend our analysis to this case in a straightforward way. As a way out we return to the Fourier series solution (\ref{F1a}) -- (\ref{F1c}) and the approximate determination of the resonance frequencies by the solution of $\det A^{(50)}=0$. From this the asymptotic form of $\omega_{res}^{(n)}(F)$ can be
obtained by inserting $\omega=\sum_{m=0}^{3} a_m F^{1-2m}$ into $\det A^{(50)}$ and setting the four highest even orders of $F$ to zero. This yields
\begin{eqnarray}
\nonumber
  \omega_{res}^{(1)}(F) &\approx& 0.415831\,F+\frac{0.87256}{F}+\frac{0.404226}{F^3}-\frac{3.83313}{F^5} \\
 &&+O(F^{-7})\;,\label{V12a}\\
 \nonumber
   \omega_{res}^{(2)}(F) &\approx&0.181157 F+\frac{0.496818}{F}+\frac{1.03437}{F^3}-\frac{12.9166}{F^5}\\
 &&+O(F^{-7})\;,\label{V12b}\\
 \nonumber
  \omega_{res}^{(3)}(F) &\approx& 0.115557 F+\frac{0.356526}{F}+\frac{1.32633}{F^3}-\frac{25.278}{F^5}\\
  &&+O(F^{-7})\;.\label{V12c}
\end{eqnarray}

The first terms proportional to $F$ are the numerical approximations of the known exact value $\frac{F}{j_{0,n}}$, but the next terms could only be determined numerically. For an alternative approach see \cite{HPS73} and \cite{AB74}.
Figure \ref{FIGQ6} shows the numerically determined resonance curves ${\mathcal R}_n$ together with the approximations
(\ref{V12a}) -- (\ref{V12c}) for $n=1,2,3$ that are valid for large $F$ and $\omega$.  Recall that according to (\ref{H6}) the limit of the unscaled quantities $F,\omega\rightarrow\infty$ is equivalent to $\widetilde{\omega_0}\rightarrow 0$ for the scaled quantity $\widetilde{\omega_0}$.

%%%%%%%%%%%%%%%%%%%%%%%%%%%%%%%%%%%%%%%%%%%%%%%%%%%%%%%%%%%%%%%%%%%%%%%%%%%%%%%%%%%%%%%%%%%%%%%%%%%%%%%%%%%%%%%%%%%%%%%%%%%%%%%
\subsubsection{The Schr\"odinger equation}\label{sec:SE}
%%%%%%%%%%%%%%%%%%%%%%%%%%%%%%%%%%%%%%%%%%%%%%%%%%%%%%%%%%%%%%%%%%%%%%%%%%%%%%%%%%%%%%%%%%%%%%%%%%%%%%%%%%%%%%%%%%%%%%%%%%%%%%%

For the sake of completeness we will show that the limit $\omega_0\rightarrow 0$ can also be considered directly for the Schr\"odinger equation
and yields an equivalent result for the linear term of the  quasienergy series w.~r.~t.~$\omega_0$.

It is convenient to consider the Hamiltonian
\begin{equation}\label{SE1}
  \hat{H}=\frac{1}{2}\left(
  \begin{array}{cc}
    F\, \cos\omega t & \omega_0\\
    \omega_0 &  -F\, \cos\omega t
  \end{array}
  \right)\;,
\end{equation}
that is unitarily equivalent to the RPL Hamiltonian hitherto considered. We make the following series ansatz for the solution of
the corresponding Schr\"odinger equation:
\begin{eqnarray}
\label{SE2a}
  \psi_1 &=& \sum_{n=0}^\infty \psi_1^{(2n)}\omega_0^{2n}\;,\\
  \label{SE2b}
  \psi_2 &=& \sum_{n=0}^\infty \psi_2^{(2n+1)}\omega_0^{2n+1}\;,
\end{eqnarray}
and obtain the following system of (in)homogeneous linear differential equations:
\begin{eqnarray}
\label{SE3a}
 {\sf i}\,\frac{d}{dt}\psi_1^{(0)} &=& \frac{F}{2}\cos\omega t\,\psi_1^{(0)}\;, \\
 \label{SE3b}
 {\sf i}\,\frac{d}{dt}\psi_2^{(2n+1)} &=& \frac{1}{2}\psi_1^{(2n)}-\frac{F}{2}\cos\omega t\,\psi_2^{(2n+1)}\;, \\
 \label{SE3c}
  {\sf i}\,\frac{d}{dt}\psi_1^{(2n+2)} &=& \frac{1}{2}\psi_2^{(2n+1)}+\frac{F}{2}\cos\omega t\,\psi_1^{(2n+2)}\;,
\end{eqnarray}
for $n=0,1,\ldots$. The two lowest terms of the series (\ref{SE2a}), (\ref{SE2b}) can be obtained in a straightforward manner:
\begin{eqnarray}
\label{SE4a}
  \psi_1^{(0)} &=& \exp\left( -{\sf i}\,\frac{F}{2\omega}\sin\omega t\right)\;, \\
  \nonumber
 \psi_2^{(1)}  &=& -\frac{{\sf i}}{2}\left(\int_{0}^{t}\exp\left( -{\sf i}\,\frac{F}{\omega}\sin\omega t'\right)dt' \right)\times\\
 \label{SE4b}
 && \exp\left({\sf i}\,\frac{F}{2\omega}\sin\omega t\right)\;.
\end{eqnarray}
We could not calculate the integral in (\ref{SE4b}) in closed form but only in form of a series using again the Jacobi-Anger expansion
and setting $f\equiv \frac{F}{\omega}$:
\begin{eqnarray}\nonumber
 && \int_{0}^{t}\exp\left( -{\sf i}\,f\,\sin\omega t'\right)dt'\\
 \nonumber
 & =&
  J_0(f)\,t \\
  \nonumber
  && +2\,{\sf i}\,\sum_{n=0}^\infty J_{2n+1}(f)\frac{\cos((2n+1)\omega t)-1}{(2n+1)\,\omega}\\
  \label{SE5}
 && +2\sum_{n=1}^\infty J_{2n}(f)\frac{\sin(2n\omega t)}{2n\,\omega}
 \;.
\end{eqnarray}
For $t=0$ we have $\psi_1^{(0)}=1$ and $\psi_2^{(1)}=0$. A second solution can be obtained that is orthogonal to
the first one such that the resulting unitary evolution operator
\begin{equation}\label{SE6}
 U(t)=\left(
 \begin{array}{cc}
   \psi_1^{(0)} & -\overline{\psi_2^{(1)}}\omega_0 \\
   \psi_2^{(1)}\omega_0 & \overline{\psi_1^{(0)}}
 \end{array}
 \right)+O(\omega_0^2)
\end{equation}
satisfies (\ref{Gen1}) with initial condition (\ref{Gen2}). The corresponding monodromy matrix
reads
\begin{equation}\label{SE7}
{\mathcal F}=U(T)=\left(
 \begin{array}{cc}
  1 & -\frac{{\sf i}\omega_0}{2}J_0(f)T \\
  -\frac{{\sf i}\omega_0}{2}J_0(f)T & 1
 \end{array}
 \right)+O(\omega_0^2)\;,
\end{equation}
and has the eigenvalues $1\pm {\sf i}\,\frac{\omega_0}{2}\,J_0(f)\,T$. This yields the quasienergies
$\epsilon=\pm \frac{\omega_0}{2}\,J_0(f)+O(\omega_0^3)$ in accordance with (\ref{V10}). One may show that
$\epsilon$ can be expanded into an odd series w.~r.~t.~$\omega_0$, whence the above term $O(\omega_0^3)$ for the next order.

%%%%%%%%%%%%%%%%%%%%%%%%%%%%%%%%%%%%%%%%%%%%%%%%%%%%%%%%%%%%%%%%%%%%%%%%%%%%%%%%%%%%%%%%%%%%%%%%%%%%%%%%%%%%%%%%%%%%%%%%%%%%%%%
\subsection{Limit case $\omega\rightarrow 0$}\label{sec:LCOM}
%%%%%%%%%%%%%%%%%%%%%%%%%%%%%%%%%%%%%%%%%%%%%%%%%%%%%%%%%%%%%%%%%%%%%%%%%%%%%%%%%%%%%%%%%%%%%%%%%%%%%%%%%%%%%%%%%%%%%%%%%%%%%%%

It is plausible that for $\omega\rightarrow 0$ the classical spin vector ${\mathbf X}(t)$ follows the magnetic field,
i.~e.~${\mathbf X}(t)=\frac{{\mathbf h}(t)}{\left\|{\mathbf h}(t)\right\|}$.
We will confirm this by calculating the Taylor series expansion of ${\mathbf X}(t)$ w.~r.~t.~$\omega$:
\begin{equation}\label{IV1}
{\mathbf X}(t)= \sum_{n=0}^{\infty}{\mathbf X}_n(t)\,\omega^n\;.
\end{equation}
Note that $\|{\mathbf X}(t)\|^2$ has the series expansion (suppressing the $t$-dependence):
\begin{eqnarray}
\nonumber
 {\mathbf X}\cdot{\mathbf X} &=&  {\mathbf X}_0\cdot{\mathbf X}_0+2\omega\, {\mathbf X}_0\cdot{\mathbf X}_1 \\
 \nonumber
  && +\omega^2\,\left(2 {\mathbf X}_0\cdot{\mathbf X}_2+ {\mathbf X}_1\cdot{\mathbf X}_1 \right) \\
  \nonumber
   && +2\omega^3\,\left( {\mathbf X}_0\cdot{\mathbf X}_3+ {\mathbf X}_1\cdot{\mathbf X}_2 \right) \\
   \nonumber
    && +\omega^4\,\left(2 {\mathbf X}_0\cdot{\mathbf X}_4+ 2{\mathbf X}_1\cdot{\mathbf X}_3+{\mathbf X}_2\cdot{\mathbf X}_2 \right) \\
    \label{IV1a}
    &&+\ldots
\end{eqnarray}
Since the normalization condition ${\mathbf X}\cdot{\mathbf X}=1$ must hold in each order of $\omega$
it follows that ${\mathbf X}_0\cdot{\mathbf X}_0=1$, but ${\mathbf X}_0\cdot{\mathbf X}_1=0$
and all other terms in the brackets of (\ref{IV1a}) have to vanish.

Since the series coefficients ${\mathbf X}_n(t)$ are $T$-periodic functions of $t$ and can be written as Fourier series each
differentiation of  ${\mathbf X}_n(t)$ w.~r.~t.~$t$ produces a factor $\omega$ and both sides of the equation of motion
\begin{equation}\label{IV2}
 \frac{d}{dt}{\mathbf X}(t)={\mathbf h}(t)\times{\mathbf X}(t)
\end{equation}
are Taylor series in $\omega$. This yields a recursive procedure to determine the ${\mathbf X}_n(t)$.

The $\omega^0$-terms of (\ref{IV2}) yield ${\mathbf 0}={\mathbf h}(t)\times{\mathbf X}_0(t)$. Together with the normalization
condition this implies (up to a sign)
\begin{equation}\label{IV3}
  {\mathbf X}_0(t)=\frac{1}{\sqrt{\omega_0^2+F^2\,\cos^2\,\omega\,t}}
   \left( \begin{array}{c}
     F\,\cos\,\omega \,t  \\
         0                \\
         \omega_0
         \end{array}
 \right)\;,
\end{equation}
which confirms the above assertion that the classical spin vector, up to normalization, follows the magnetic field.

The next order, linear in $\omega$, yields
\begin{equation}\label{IV4}
 \frac{d}{dt}{\mathbf X}_0(t)=\omega\,{\mathbf h}(t)\times{\mathbf X}_1(t)\;.
\end{equation}
This is an inhomogeneous linear equation with the general solution
\begin{equation}\label{IV5}
\omega\, {\mathbf X}_1(t)=\frac{d}{dt}{\mathbf X}_0(t)\times\frac{{\mathbf h}(t)}{\left\|{\mathbf h}(t)\right\|^2}
 +\omega\,\lambda_1(t)\,\mathbf{h(t)}\;.
\end{equation}
The normalization condition implies ${\mathbf X}_0\cdot{\mathbf X}_1 =0$ and hence $\lambda_1(t)=0$. It follows that
\begin{equation}\label{IV5a}
{\mathbf X}_1(t)=\left(0,\frac{F  \omega_0 \sin (\omega\,t)}{\left(F^2 \cos ^2(\omega\,t)+ \omega_0^2\right)^{3/2}},0\right)^\top\;.
\end{equation}

In the next quadratic order of $\omega$ we
analogously have
\begin{equation}\label{IV6}
 \frac{d}{dt}{\mathbf X}_1(t)=\omega\,{\mathbf h}(t)\times{\mathbf X}_2(t)\;,
\end{equation}
with the general solution
\begin{equation}\label{IV7}
\omega\, {\mathbf X}_2(t)=\frac{d}{dt}{\mathbf X}_1(t)\times\frac{{\mathbf h}(t)}{\left\|{\mathbf h}(t)\right\|^2}
 +\omega\,\lambda_2(t)\,\mathbf{h(t)}\;.
\end{equation}

This time the  normalization condition up to the quadratic order of $\omega$ gives
$2 {\mathbf X}_0\cdot{\mathbf X}_2+ {\mathbf X}_1\cdot{\mathbf X}_1 ={\mathbf 0}$ and hence
\begin{equation}\label{IV8}
 \lambda_2(t)=-\frac{1}{2\left\|{\mathbf h}\right\|}{\mathbf X}_1\cdot{\mathbf X}_1=
 -\frac{F^2 \omega_0 ^2 \sin ^2(\omega\,t)}{2 \left(F^2 \cos ^2(\omega\,t)+\omega_0 ^2\right)^{7/2}}\;.
\end{equation}
The corresponding ${\mathbf X}_2(t)$ will not be displayed here.

In this way we may recursively determine an arbitrary number of terms ${\mathbf X}_n(t)$.
It can be shown by induction over $n$ that for odd $n$ the ${\mathbf X}_n(t)$ have only a
non-vanishing $y$-component and for even $n$ the $y$-component vanishes.
Hence ${\mathbf X}_n(t)\cdot{\mathbf X}_m(t)=0$ if $n$ and $m$ have different parity.
This implies that the pre-factors of $\omega^n$ in (\ref{IV1a}) vanish  and hence $\lambda_n(t)=0$ for all odd $n$.

We note that the Taylor expansion (\ref{IV1}) breaks down for $\omega_0\rightarrow 0$. This follows already from
the observation that the velocity $\left\| \frac{d}{dt}\mathbf{X}_0(t)\right\|$ would assume arbitrary large values for $\omega_0\rightarrow 0$
if (\ref{IV1}) would be a correct description of the solution $\mathbf{X}(t)$.
However, the velocity is bounded by $\left\|{\mathbf h} \right\|$ and hence (\ref{IV1}) cannot longer hold. For example, we may consider a small fixed value of $\omega$ and a finite value of $\omega_0$ such that $\mathbf{X}_0(t)$ is a good approximation of the exact solution $\mathbf{X}(t)$. If we lower $\omega_0$ to smaller and smaller values we would obtain a sudden switch to the behavior described in section \ref{sec:LCOM0} for the limit $\omega_0\rightarrow 0$. In this sense the two limits $\omega\rightarrow 0$ and $\omega_0\rightarrow 0$ cannot be interchanged.

Finally, we will consider the quasienergy  for the lowest orders of $\omega$.
In the limit  $\omega\rightarrow 0$ the geometrical part of the quasienergy vanishes since the solution ${\mathbf X}_0(t)$ is confined to the $x-z$-plane.
The dynamical part (\ref{C25c}) has the value
\begin{eqnarray}\label{F8a}
\epsilon_d=\epsilon&=&\overline{\frac{1}{2R}\left( F\,\cos\omega t\, X + \omega_0 Z \right)}\\
\label{F8b}
&=&\overline{\left(\frac{\left(\omega_0^2+F^2\,\cos^2\omega t \right)}{2\sqrt{\omega_0^2+F^2\cos^2\omega t}}\right)}\\
\label{F8c}
&=&\frac{\omega}{4\pi}\int_{0}^{2\pi/\omega} \sqrt{\omega_0^2+F^2\cos^2\omega t}\,dt \\
\label{F8d}
&=&\frac{\omega_0}{\pi }\, E\left(-\frac{F^2}{\omega_0^2}\right)\;,
\end{eqnarray}
where $E(\ldots)$ denotes the complete elliptic integral of the second kind. See also \cite{S65} for a similar result.
In the special case $F=1/2$ and $\omega_0=1$ that is portrayed in Figure \ref{FIGQ4} we have
\begin{equation}\label{F9}
\epsilon\rightarrow \epsilon_0\equiv \frac{E\left(-\frac{1}{4}\right)}{\pi }
\approx 0.52992
\end{equation}
for $\omega\rightarrow 0$.

\begin{figure}[ht]
  \centering
    \includegraphics[width=1.0\linewidth]{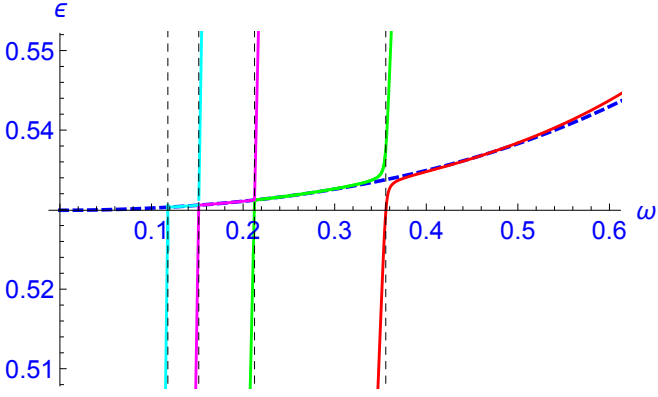}
  \caption[BS1]
  {A detail of Figure \ref{FIGQ4} showing various branches of the quasienery (continuous colored curves),
  the corresponding  resonance frequencies $\omega_{res}^{(n)}$ (vertical dashed lines) and the asymptotic limit
  $\epsilon_{asy}=\epsilon_0+\epsilon_2\omega^2+\epsilon_4\omega^4$ (blue dashed curve) for $\omega\rightarrow 0$ according to (\ref{F9}), (\ref{F11}) and (\ref{F12}).
  }
  \label{FIGQ4a}
\end{figure}

The next corrections to $\epsilon$ are of the form $\epsilon_2\,\omega^2+\epsilon_4\,\omega^4$. We obtain
\begin{equation}\label{F10}
\epsilon_2=
\frac{\left(2 F^2+\omega_0^2\right) E\left(\frac{F^2}{F^2+\omega_0^2}\right)-
\omega_0^2 K\left(\frac{F^2}{F^2+\omega_0^2}\right)}{6 \pi \omega_0^2 \sqrt{F^2+\omega_0^2}}\;,
\end{equation}
and
\begin{eqnarray}
\nonumber
  \epsilon_4 &=& \frac{1}{60\, \pi \, \omega _0^6\, \left(F^2+\omega _0^2\right){}^{5/2}} \\
  \nonumber
  && \left[  \frac{1}{6} \left(64 F^8+200 F^6 \omega _0^2+231 F^4 \omega _0^4+137 F^2 \omega_0^6\right.\right.\\
  \nonumber
  && \left.- 14\, \omega _0^8\right) E\left(\frac{F^2}{F^2+\omega _0^2}\right)
  \\
 \nonumber
   && -\frac{1}{3} \omega _0^2 \left(16 F^6+36 F^4 \omega _0^2+27 F^2 \omega _0^4-7\, \omega_0^6\right) \\
   \label{F10a}
   &&\left.
   K\left(\frac{F^2}{F^2+\omega _0^2}\right)
   \right]
   \;,
\end{eqnarray}
where $K(\ldots)$ and $E(\ldots)$ denote the complete elliptic integrals of the first and second kind, resp.~.

It is plausible that the asymptotic limit $\epsilon_{asy}$ of $\epsilon$
for $\omega\rightarrow 0$ does not approximate a single branch of the quasienergy but rather represents a kind of envelope of the various branches, see Figure \ref{FIGQ4a}.
In the special case $F=1/2$ and $\omega_0=1$ that is portrayed in Figure \ref{FIGQ4a} we have
\begin{equation}\label{F11}
\epsilon_2=\frac{3 E\left(\frac{1}{5}\right)-2 \,K\left(\frac{1}{5}\right)}{6 \sqrt{5}\pi }
\approx 0.0272334\;,
\end{equation}
and
\begin{equation}\label{F12}
\epsilon_4=\frac{203 E\left(\frac{1}{5}\right)-24 K\left(\frac{1}{5}\right)}{1500 \sqrt{5} \pi }
\approx0.0249063\;.
\end{equation}

Since $\epsilon_{asy}$ represents the envelope of the branches of $\epsilon$ the asymptotic form of the resonance frequencies cannot
be determined by the present method. However, the inspection of Figure \ref{FIGG4} suggests that for $\omega\rightarrow 0$ the resonance frequencies
are given by an interpolation between the limits for $F\rightarrow 0$ and $\omega_0\rightarrow 0$, namely
\begin{equation}\label{F13}
 \omega_{res}^{(n)} \sim \frac{F}{j_{0,n}}+\frac{\omega _0}{2 n-1}
 \;.
\end{equation}
This approximation is of reasonable quality for small $F$ or small $\omega_0$ but of poor quality for $F\sim\omega_0$
since there the small curvature of the resonance curves ${\mathcal R}_n$ in the triangular domain $\triangle$, see Figure \ref{FIGG4},
should be taken into account.

%%%%%%%%%%%%%%%%%%%%%%%%%%%%%%%%%%%%%%%%%%%%%%%%%%%%%%%%%%%%%%%%%%%%%%%%%%%%%%%%%%%%%%%%%%%%%%%%%%%%%%%%%%%%%%%%%%%%%%%%%%%%%%%
\subsection{Limit case $\omega \rightarrow \infty$}\label{sec:FQW}
%%%%%%%%%%%%%%%%%%%%%%%%%%%%%%%%%%%%%%%%%%%%%%%%%%%%%%%%%%%%%%%%%%%%%%%%%%%%%%%%%%%%%%%%%%%%%%%%%%%%%%%%%%%%%%%%%%%%%%%%%%%%%%

To investigate the limit $\omega\rightarrow\infty$ we set ${\mathcal T}\equiv \frac{1}{\omega}$ and make the following ansatz of
an FT series:
\begin{eqnarray}
\label{W1a}
X(t) &=& \sum_{n=2,4,\ldots}^{\infty}{\mathcal T}^n\sum_{m=1,3,\ldots}^{n-1}x_{n,m}\,\cos(m\omega t)\;, \\
\label{W1b}
Y(t) &=& \sum_{n=1,3,\ldots}^{\infty}{\mathcal T}^n\sum_{m=1,3,\ldots}^{n}y_{n,m}\,\sin(m\omega t)\;, \\
\label{W1c}
Z(t) &=& 1+\sum_{n=2,4,\ldots}^{\infty}{\mathcal T}^n\sum_{m=2,4,\ldots}^{n}z_{n,m}\,\cos(m\omega t).
\end{eqnarray}

This ansatz is inserted into the classical equations of motion (\ref{AA1a}) -- (\ref{AA1c}) in such a way that each factor
$\omega$ resulting from the differentiation $\frac{d}{dt}$ is replaced by $1/{\mathcal T}$. As usual, the condition that the resulting FT series has vanishing coefficients yields linear equations that determine the $x_{n,m},\,y_{n,m}$ and $z_{n,m}$ and hence $X(t),\,Y(t)$ and $Z(t)$ up to any finite order.
In lowest non-trivial order the asymptotic form of the solution reads (re-substituting ${\mathcal T}=1/\omega$):
\begin{eqnarray}
\label{W2a}
X(t) &=& -\frac{F \omega_0} {\omega ^2}\,\cos (\omega\,t )+O(\omega^{-4})=\frac{X(t)}{R}\;, \\
\label{W2b}
Y(t) &=& -\frac{F} {\omega }\,\sin (\omega\,t )+O(\omega^{-3})=\frac{Y(t)}{R}\;, \\
\label{W2c}
Z(t) &=& 1+\frac{F^2} {4\,\omega ^2}\,\cos (2\,\omega\,t )+O(\omega^{-4})\;,\\
\label{W2d}
\frac{Z(t)}{R}&=&1-\frac{F^2 (1-\cos (2 \omega t ))}{4\, \omega ^2}+O(\omega^{-4})\;.
\end{eqnarray}

We will compare this result with the first terms of the $1/\omega$-Taylor expansion of the normalized classical RPC solution ${\mathbf X}_-(t)$
according to (\ref{C20}):
\begin{eqnarray}
\label{W3a}
X(t) &=&-\left(\frac{F } {\omega} +\frac{F \omega_0} {\omega ^2}\right)\cos (\omega\,t )+O(\omega^{-3})\;, \\
\label{W3b}
Y(t) &=&-\left(\frac{F}  {\omega} +\frac{F \omega_0} {\omega ^2}\right)\,\sin (\omega\,t )+O(\omega^{-3})\;, \\
\label{W3c}
Z(t) &=& 1-\frac{F^2} {2\,\omega ^2}+O(\omega^{-3})\;.
\end{eqnarray}

Despite some similarities we come to the conclusion that both solutions are different, even in the lowest non-vanishing
order w.~r.~t.~$1/\omega$. This is in contrast to the view that the rotating wave approximation is an analytical approximation to the RPL solution
that is asymptotically valid in the limit of large $\omega$.

According to the FT solution the quasienergy $\epsilon(\omega_0,F,\omega)$ can be calculated as a power series in $1/\omega$
the first terms of which are:
\begin{equation}\label{W4}
\epsilon(\omega_0,F,\omega)=\frac{\omega_0}{2}-\frac{F^2 \omega_0}{8 \omega ^2}+\frac{F^2 \omega_0 \left(F^2-16 \omega_0^2\right)}{
 128 \omega^4}+O(\omega^{-6})\;.
\end{equation}

This is in accordance with the series expansion of (\ref{V10})
\begin{equation}\label{W5}
\frac{\omega_0}{2}J_0\left(\frac{F}{\omega}\right)=
\frac{\omega_0}{2}-\frac{F^2 \omega_0}{8 \omega ^2}+\frac{F^4 \omega_0 }{128 \omega^4}+O(\omega^{-6})\;,
\end{equation}
keeping in mind that (\ref{V10}) holds only in first order in $\omega_0$.

%%%%%%%%%%%%%%%%%%%%%%%%%%%%%%%%%%%%%%%%%%%%%%%%%%%%%%%%%%%%%%%%%%%%%%%%%%%%%%%%%%%%%%%%%%%%%%%%%%%%%%%%%%%%%%%%%%%%%%%%%%%%%%%
\subsection{Limit case $\omega_0 \rightarrow \infty$}\label{sec:FQX}
%%%%%%%%%%%%%%%%%%%%%%%%%%%%%%%%%%%%%%%%%%%%%%%%%%%%%%%%%%%%%%%%%%%%%%%%%%%%%%%%%%%%%%%%%%%%%%%%%%%%%%%%%%%%%%%%%%%%%%%%%%%%%%

As remarked above, due to (\ref{H6}) this limit is equivalent to the limit $\widetilde{F}\rightarrow 0,\,\widetilde{\omega}\rightarrow 0$
and $\widetilde{\omega_0}\rightarrow 1$ of the scaled quantities. First we will compare the limit of ${\mathbf X}(t)$ for $\omega\rightarrow 0$
according to (\ref{IV3}) and  (\ref{IV5a}) with FT series expansion (\ref{FTX1a}), (\ref{FTX1b})  of ${\mathbf X}(t)$ that holds for $F\rightarrow 0$.
Note that for the comparison the latter one has to be normalized. We obtain the result that both limits coincide if we ignore terms of the order
$O(F^3)$ and $O(\omega^2)$:
\begin{eqnarray}
\label{FQX1a}
  X(t) &=& \frac{F \cos (\omega t  )}{\omega _0}+ O(F^3,\omega^2)\;, \\
 Y(t) &=& \frac{F \omega  \sin (\omega t)}{\omega _0^2} + O(F^3,\omega^2)\;, \\
  Z(t) &=& 1-\frac{F^2 (1+\cos 2 \omega t)}{4 \omega _0^2} + O(F^3,\omega^2)\;.
\end{eqnarray}
In deriving this result we used, of course, a restricted series expansion w.~r.~t.~$\omega$ that leaves the terms $\cos n\omega t$ and $\sin m\omega t$
of the Fourier series intact.

Analogously, we will compare the asymptotic forms of the quasienergy for $\omega\rightarrow 0$ according to (\ref{F8d})  and for $F\rightarrow 0$
according to (\ref{FQ2}). Again, we find that both limits are compatible and yield the common result:
\begin{equation}\label{FQX2}
 \epsilon= \frac{\omega _0}{2}+\frac{F^2}{8 \omega _0}+ O(F^3,\omega^2)\;.
\end{equation}

%%%%%%%%%%%%%%%%%%%%%%%%%%%%%%%%%%%%%%%%%%%%%%%%%%%%%%%%%%%%%%%%%%%%%%%%%%%%%%%%%%%%%%%%%%%%%%%%%%%%%%%%%%%%%%%%%%%%%%%%%%%%%%%
\section{Summary and Outlook}\label{sec:SO}
%%%%%%%%%%%%%%%%%%%%%%%%%%%%%%%%%%%%%%%%%%%%%%%%%%%%%%%%%%%%%%%%%%%%%%%%%%%%%%%%%%%%%%%%%%%%%%%%%%%%%%%%%%%%%%%%%%%%%%%%%%%%%%%

We have revisited the Floquet theory of two level systems and suggested a kind of geometrical approach
based on periodic solutions of the classical equation of motion that can be visualized by closed trajectories on the Bloch sphere.
From these solutions one can reconstruct the Floquet solutions of the underlying Schr\"odinger equation
including the quasienergy $\epsilon$ by calculating the coefficients of a Fourier series.
The relation of $\epsilon$ to the classical action integral and the splitting of the quasienergy into a geometrical and a dynamical part,
$\epsilon=\epsilon_g +\epsilon_d$, fits well into this geometrical setting.
In the case of the RPE the partial derivatives of $\epsilon$ w.~r.~t.~the system's parameters
$\omega_0,F$ and $\omega$ can be calculated by methods of analytical mechanics. Thereby the resonance condition of J.~H.~Shirley
receives a geometrical/dynamical interpretation and the noteworthy relation
$\frac{\partial \epsilon}{\partial \omega}=\frac{\epsilon_g}{\omega}$  is derived that holds generally for two level systems.

The mentioned results are proven not with strict mathematical rigor, but according to the usual standards of theoretical physics.
This means that there are, besides the technical subtleties, still minor logical gaps. For example,
it would be desirable to clarify the validity of the assumptions of Assertion \ref{ass2} on the existence of exactly
two normalized periodic solutions of the classical equation of motion. Another interesting open problem is the proof of the continuity
or even analyticity of the quasienergy as a function of one of the parameters $\omega_0,F$ and $\omega$.
As briefly mentioned in section \ref{sec:GenSU2}
the quasienergy can be viewed as an eigenvalue of the Floquet Hamiltonian defined on an extended Hilbert space. Hence one might
invoke the corresponding theory of analytical perturbations, e.~g.~, Rellich's theorem \cite{S91} or similar tools,
but it is not clear whether the Floquet Hamiltonian satisfies the pertaining conditions.

We have checked our results for simple solvable examples, but the main intended application is the RPL case.
Here our approach leads to certain analytical approximations that can be conveniently handled by computer-algebraic aids.
It is also possible to perform the geometrical approach for the various limit cases of the RPL. We have compared these results
with those known from the literature only in a few cases, since a thorough comparison would need too much space, but such a comparison
is nevertheless desirable.

Another future task would be the attempt to utilize the geometrical approach to obtain examples of the theory of periodic thermodynamics, that
describe  periodically driven two level systems coupled to a heat bath. For recent approaches to this problem,
see, e.~g.~, \cite{YLZ15}, \cite{LH14} and \cite{AGK12} -- \cite{SMM18}.

%%%%%%%%%%%%%%%%%%%%%%%%%%%%%%%%%%%%%%%%%%%%%%%%%%%%%%%%%%%%%%%%%%%%%%%%%%%%%%%%%%%%%%%%%%%%%%%%%%%%%%%%%%%%%%%%%%%%%%%%%%%%%%%%%%%%%%%%%%
\section*{Acknowledgment}
%%%%%%%%%%%%%%%%%%%%%%%%%%%%%%%%%%%%%%%%%%%%%%%%%%%%%%%%%%%%%%%%%%%%%%%%%%%%%%%%%%%%%%%%%%%%%%%%%%%%%%%%%%%%%%%%%%%%%%%%%%%%%%%%%%%%%%%%%

I am indebted to the members of the DFG research group FOR 2692 for continuous support and encouragement, especially to Martin Holthaus and J\"urgen Schnack.  Moreover, I gratefully acknowledge discussions with Thomas Br\"ocker on the subject of this paper.
I dedicate this work to my friend and mentor Marshall Luban on the occasion of his $82$th birthday.

\end{document}